# Massive MIMO Performance With Imperfect Channel Reciprocity and Channel Estimation Error

De Mi, Mehrdad Dianati, Lei Zhang, Sami Muhaidat, *Senior Member, IEEE*, and Rahim Tafazolli

*Abstract*—Channel reciprocity in time-division duplexing (TDD) massive multiple-input multiple-output (MIMO) systems can be exploited to reduce the overhead required for the acquisition of channel state information (CSI). However, perfect reciprocity is unrealistic in practical systems due to random radio-frequency (RF) circuit mismatches in uplink and downlink channels. This can result in a significant degradation in the performance of linear precoding schemes, which are sensitive to the accuracy of the CSI. In this paper, we model and analyse the impact of RF mismatches on the performance of linear precoding in a TDD multi-user massive MIMO system, by taking the channel estimation error into considerations. We use the truncated Gaussian distribution to model the RF mismatch, and derive closed-form expressions of the output signal-to-interference-plus-noise ratio for maximum ratio transmission and zero forcing precoders. We further investigate the asymptotic performance of the derived expressions, to provide valuable insights into the practical system designs, including useful guidelines for the selection of the effective precoding schemes. Simulation results are presented to demonstrate the validity and accuracy of the proposed analytical results.

## I. INTRODUCTION

MASSIVE (or large scale) MIMO (multiple-input multiple-output) systems have been identified as enabling technologies for the 5th Generation (5G) of wireless systems [1]–[5]. Such systems propose the use of a large number of antennas at the base station (BS) side. A notable advantage of this approach is that it allows the use of simple processing at both uplink (UL) and downlink (DL)

Manuscript received August 9, 2016; revised December 5, 2016; accepted February 17, 2017. Date of publication March 1, 2017; date of current version September 14, 2017. The authors would like to acknowledge the support of the University of Surrey 5GIC (http://www.surrey.ac.uk/5gic) members for this work. This work has also been supported by the European Union Seventh Framework Programme (FP7/2007-2013) under Grant 619563 (MiWaveS). The associate editor coordinating the review of this paper and approving it for publication was X. Yuan. *(Corresponding author: Sami Muhaidat.)*

D. Mi, L. Zhang, and R. Tafazolli are with the 5G Innovation Centre, Institute for Communication Systems, University of Surrey, Guildford GU2 7XH, U.K. (e-mail: d.mi@surrey.ac.uk; lei.zhang@surrey.ac.uk; r.tafazollig@surrey.ac.uk).

M. Dianati is with the Warwick Manufacturing Group, University of Warwick, Coventry CV4 7AL, U.K., and also with the 5G Innovation Centre, Institute for Communication Systems, University of Surrey, Guildford GU2 7XH, U.K. (e-mail: m.dianati@warwick.ac.uk).

S. Muhaidat is with the Department of Electrical and Computer Engineering, Khalifa University, Abu Dhabi 127788, United Arab Emirates, and also with the 5G Innovation Centre, Institute for Communication Systems, University of Surrey, Guildford GU2 7XH, U.K. (e-mail: muhaidat@ieee.org).

directions [6], [7]. For example, for the DL transmission, two commonly known linear precoding schemes, i.e., *maximum ratio transmission* (MRT) and *zero-forcing* (ZF), have been extensively investigated in the context of massive MIMO systems [8]–[10]. It has been shown that both schemes perform well with a relatively low computational complexity [8], and can achieve a spectrum efficiency close to the optimal non-linear precoding techniques, such as dirty paper coding [9], [11]. However, the price to pay for the use of simple linear precoding schemes is the overhead required for acquiring the instantaneous channel state information (CSI) in the massive MIMO systems [10], [12].

In principle, massive MIMO can be adopted in both frequency-division duplexing (FDD) and time-division duplexing (TDD) systems. Nevertheless, the overhead of CSI acquisition in FDD massive MIMO systems is considerably higher than that of TDD systems, due to the need for a dedicated feedback channel and the infeasible number of pilots, which is proportional to the number of BS antennas [13]. On the contrary, by exploiting the channel reciprocity in TDD systems, the BS can estimate the DL channel by using the UL pilots from the user terminals (UTs). Hence, there is no feedback channel required, and the overhead of the pilot transmission is proportional to the number of UTs antennas, which is typically much less than the number of BS antennas in massive MIMO systems [9]. Therefore, TDD operation has been widely considered in the system with large-scale antenna arrays [1], [7]–[9].

Most prior studies assume perfect channel reciprocity by constraining that the time delay from the UL channel estimation to the DL transmission is less than the coherence time of the channel [1], [7], [8]. Such an assumption ignores two key facts: 1) UL and DL radio-frequency (RF) chains are separate circuits with random impacts on the transmitted and received signals [2], [6]; 2) the interference profile at the BS and UT sides may be significantly different [14]. The former phenomenon is known as *RF mismatch* [15], which is the main focus of this paper. RF mismatches can cause random deviations of the estimated values of the UL channel from the actual values of the DL channel within the coherent time of the channel. Such deviations are known as reciprocity errors that invalidate the assumption of perfect reciprocity.

The existing works on studying reciprocity errors can be divided into two categories. In the first category, e.g. [16], reciprocity errors are considered as an additive random uncertainty to the channel coefficients. However, it is shown in [15] that additive modelling of the reciprocity errors is inadequate in capturing the full impact of RF mismatches. Therefore, the recent works consider multiplicative reciprocity errors where the channel coefficients are multiplied by random complex numbers representing the reciprocity errors. For example, the works in [17] and [18] model the reciprocity errors as uniformly distributed random variables which are multiplied by the channel coefficients. The authors model the amplitude and phase of the multiplicative reciprocity error by two independent and uniformly distributed random variables, i.e., amplitude and phase errors. Rogalin et al. in [17] propose a calibration scheme to deal with reciprocity errors. Zhang et al. in [18] propose an analysis of the performance of MRT and regularised ZF precoding schemes. Practical studies [19]–[21] argue that the use of uniform distributions for modelling phase and amplitude errors is not realistic. Alternatively, they suggest the use of *truncated Gaussian distributions* instead. However, these works do not provide an in-depth analysis of the impact of reciprocity errors. In this paper, we aim to fill this research gap and present an in depth analysis of the impact of the multiplicative reciprocity errors for TDD massive MIMO systems. In addition, we also take the additive channel estimation error into considerations. The contributions of this paper can be summarised as follows:

- Under the assumption of a large number of antennas at BS and imperfect channel estimation, we derive closed-form expressions of the output SINR for ZF and MRT precoding schemes in the presence of reciprocity errors.
- We further investigate the impact of reciprocity errors on the performance of MRT and ZF precoding schemes and demonstrate that such errors can reduce the output SINR by more than 10-fold. Note that all of the analysis is considered in the presence of the channel estimation error, to show the compound effects on the system performance of the additive and multiplicative errors.
- We quantify and compare the performance loss of both ZF and MRT analytically, and provide insights to guide the choice of the precoding schemes for massive MIMO systems in the presence of the reciprocity error and estimation error.

The rest of the paper is organised as follows. In Section II, we describe the TDD massive MIMO system model with imperfect channel estimation and the reciprocity error model due to the RF mismatches. The derivations of the output SINR for MRT and ZF precoding schemes are given in Sections III. In Section IV, we analyse the effect of reciprocity errors on the output SINR when the number of BS antennas approaches infinity. Simulation results and conclusions are provided in Section V and Section VI respectively. Some of the detailed derivations are given in the appendices.

*Notations:* $\mathbb{E}\{\cdot\}$ denotes the expectation operator, and $\text{var}(\cdot)$ is the mathematical variance. Vectors and matrices are denoted by boldface lower-case and upper-case characters, and the operators $(\cdot)^*$, $(\cdot)^T$ and $(\cdot)^H$ represent complex conjugate, transpose and conjugate transpose, respectively. The $M \times M$ identity matrix is denoted by $\mathbf{I}_M$, and $\text{diag}(\cdot)$ stands for the diagonalisation operator to transform a vector to a diagonal matrix. $\text{tr}(\cdot)$ denotes the matrix trace operation. $|\cdot|$ denotes the magnitude of a complex number, while $\|\cdot\|$ is the Frobenius norm of a matrix. The imaginary unit is denoted $j$, and $\Re(\cdot)$ is the real part of a complex number. "$\triangleq$" is the equal by definition sign. The exponential function and the Gauss error function are defined as $\exp(\cdot)$ and $\text{erf}(\cdot)$, respectively.

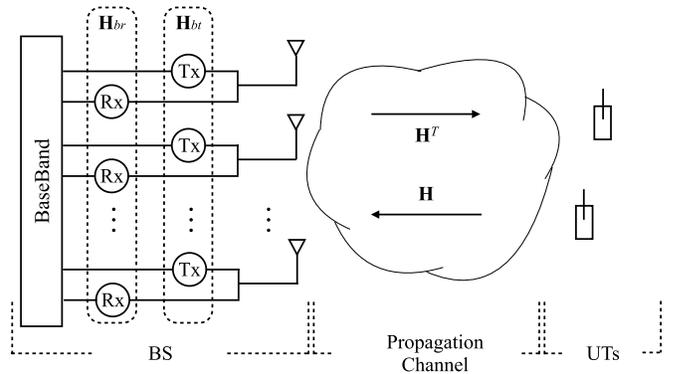

Fig. 1. A massive MU-MIMO TDD system.

## II. SYSTEM MODEL

We consider a Multi User (MU) MIMO system as shown in Fig. 1 that operates in TDD mode. This system comprises of $K$ single-antenna UTs and one BS with $M$ antennas, where $M \gg K$. Each antenna element is connected to an independent RF chain. We assume that the effect of antenna coupling is negligible, and that the UL channel estimation and the DL transmission are performed within the coherent time of the channel. In the rest of this section, we model the reciprocity errors caused by RF mismatches first, and then present the considered system model in the presence of the reciprocity error.

### A. Channel Reciprocity Error Modelling

Due to the fact that the imperfection of the channel reciprocity at the single-antenna UT side has a trivial impact on the system performance [2], we focus on the reciprocity errors at the BS side.[1] Hence, as shown in Fig. 1, the overall transmission channel consists of the physical propagation channel as well as transmit (Tx) and receive (Rx) RF frontends at the BS side. In particular, considering the reciprocity of the propagation channel in TDD systems, the UL and DL channel matrices are denoted by $\mathbf{H} \in \mathbb{C}^{M \times K}$ and $\mathbf{H}^T$, respectively. $\mathbf{H}_{br}$ and $\mathbf{H}_{bt}$ represent the effective response matrices of the Rx and Tx RF frontends at the BS, respectively. Unless otherwise stated, subscript '$b$' stands for BS, and '$t$' and '$r$' correspond to Tx and Rx frontends, respectively. $\mathbf{H}_{br}$ and $\mathbf{H}_{bt}$ can be modelled as $M \times M$ diagonal matrices, e.g., $\mathbf{H}_{br}$ can be given as

$$\mathbf{H}_{br} = \text{diag}(h_{br,1}, \cdots, h_{br,i}, \cdots, h_{br,M}), \quad (1)$$

with the $i$-th diagonal entry $h_{br,i}$, $i = 1, 2, \cdots, M$, represents the per-antenna response of the Rx RF frontend. Considering that the power amplitude attenuation and the phase shift for each RF frontend are independent, $h_{br,i}$ can be expressed as [15], [22]

$$h_{br,i} = A_{br,i} \exp(j\varphi_{br,i}), \quad (2)$$

[1] The effective responses of Tx/Rx RF frontend at UTs are set to be ones.

where $A$ and $\varphi$ denote amplitude and phase RF responses, respectively. Similarly, $M \times M$ diagonal matrix $\mathbf{H}_{bt}$ can be denoted as

$$\mathbf{H}_{bt} = \mathrm{diag}(h_{bt,1}, \cdots, h_{bt,i}, \cdots, h_{bt,M}), \quad (3)$$

with $i$-th diagonal entry $h_{bt,i}$ given by

$$h_{bt,i} = A_{bt,i}\exp(j\varphi_{bt,i}). \quad (4)$$

In practice, there might be differences between the Tx front and the Rx front in terms of RF responses. We define the RF mismatch between the Tx and Rx frontends at the BS by calculating the ratio of $\mathbf{H}_{bt}$ to $\mathbf{H}_{br}$, i.e.,

$$\mathbf{E} \triangleq \mathbf{H}_{bt}\mathbf{H}_{br}^{-1} = \mathrm{diag}(\frac{h_{bt,1}}{h_{br,1}}, \cdots, \frac{h_{bt,i}}{h_{br,i}}, \cdots, \frac{h_{bt,M}}{h_{br,M}}), \quad (5)$$

where the $M \times M$ diagonal matrix $\mathbf{E}$ can be regarded as the compound RF mismatch error, in the sense that $\mathbf{E}$ combines $\mathbf{H}_{bt}$ and $\mathbf{H}_{br}$. In (5), the minimum requirement to achieve the perfect channel reciprocity is $\mathbf{E} = c\mathbf{I}_M$ with a scalar[2] $c \in \mathbb{C}_{\neq 0}$. The scalar $c$ does not change the direction of the precoding beamformer [15], hence no impact on MIMO performance. Contrary to the case of the perfect reciprocity, in realistic scenarios, the diagonal entries of $\mathbf{E}$ may be different from each other, which introduces the RF mismatch caused channel reciprocity errors into the system. Particularly, considering the case with the hardware uncertainty of the RF frontends caused by the various of environmental factors as discussed in [2], [16], and [23], the entries become independent random variables. However, in practice, the response of RF hardware components at the Tx front is likely to be independent of that at the Rx front, which cannot be accurately represented by the compound error model $\mathbf{E}$ in (5). Hence, the separate modelling for $\mathbf{H}_{bt}$ and $\mathbf{H}_{br}$ is more accurate from a practical point of view. Therefore, we focus our investigation in this work on the RF mismatch caused reciprocity error by considering this separate error model.

Next we model the independent random variables $A_{br,i}$, $\varphi_{br,i}$, $A_{bt,i}$ and $\varphi_{bt,i}$ in (2) and (4) to reflect the randomness of the hardware components of the Rx and Tx RF frontends. Here, in order to capture the aggregated effect of the mismatch on the system performance, the phase and amplitude errors can be modelled by the truncated Gaussian distribution [20], [21], which is more generalised and realistic comparing to the uniformly distributed error model in [17] and [18]. The preliminaries of the truncated Gaussian distribution are briefly presented in Appendix A, and accordingly the amplitude and phase reciprocity errors of the Tx front $A_{bt,i}$, $\varphi_{bt,i}$ and the Rx front $A_{br,i}$, $\varphi_{br,i}$ can be modelled as

$$A_{bt,i} \sim \mathcal{N}_T(\alpha_{bt,0}, \sigma_{bt}^2), A_{bt,i} \in [a_t, b_t], \quad (6)$$

$$\varphi_{bt,i} \sim \mathcal{N}_T(\theta_{bt,0}, \sigma_{\varphi_t}^2), \varphi_{bt,i} \in [\theta_{t,1}, \theta_{t,2}], \quad (7)$$

$$A_{br,i} \sim \mathcal{N}_T(\alpha_{br,0}, \sigma_{br}^2), A_{br,i} \in [a_r, b_r], \quad (8)$$

$$\varphi_{br,i} \sim \mathcal{N}_T(\theta_{br,0}, \sigma_{\varphi_r}^2), \varphi_{br,i} \in [\theta_{r,1}, \theta_{r,2}], \quad (9)$$

where, without loss of generality, the statistical magnitudes of these truncated Gaussian distributed variables are assumed to be static, e.g., $\alpha_{bt,0}$, $\sigma_{bt}^2$, $a_t$ and $b_t$ of $A_{bt,i}$ in (6) remain constant within the considered coherence time of the channel. Notice that the truncated Gaussian distributed phase error in (7) and (9) becomes a part of exponential functions in (2) and (4), whose expectations can not be obtained easily. Thus, we provide a generic result for these expectations in the following Proposition 1.

[2]Particularly, the case with $\mathbf{E} = \mathbf{I}_M$ is equivalent to that with $\mathbf{H}_{bt} = \mathbf{H}_{br}$, which means that the Tx/Rx RF frontends have the identical responses.

*Proposition 1:* Given $x \sim \mathcal{N}_T(\mu, \sigma^2), x \in [a, b]$, and the probability density function $f(x, \mu, \sigma; a, b)$ as (59) in Appendix A. Then the mathematical expectation of $\exp(jx)$ can be expressed as

$$\mathbb{E}\{\exp(jx)\}$$
$$= \exp\left(-\frac{\sigma^2}{2} + j\mu\right)$$
$$\times \left(\frac{\mathrm{erf}\left(\left(\frac{b-\mu}{\sqrt{2\sigma^2}}\right) - j\frac{\sigma}{\sqrt{2}}\right) - \mathrm{erf}\left(\left(\frac{a-\mu}{\sqrt{2\sigma^2}}\right) - j\frac{\sigma}{\sqrt{2}}\right)}{\mathrm{erf}\left(\frac{b-\mu}{\sqrt{2\sigma^2}}\right) - \mathrm{erf}\left(\frac{a-\mu}{\sqrt{2\sigma^2}}\right)}\right). \quad (10)$$

*Proof:* See Appendix B. □

Then the phase-error-related parameters $g_t \triangleq \mathbb{E}\{\exp(j\varphi_{bt,i})\}$ and $g_r \triangleq \mathbb{E}\{\exp(j\varphi_{br,i})\}$ can be given in Appendix C by specialising Proposition 1. Also, based on (6), (8) and Appendix A, the amplitude-error-related parameters $\mathbb{E}\{A_{bt,i}\}$, $\mathbb{E}\{A_{br,i}\}$, $\mathrm{var}(A_{bt,i})$ and $\mathrm{var}(A_{br,i})$ can be given by $\alpha_t$, $\alpha_r$, $\sigma_t^2$ and $\sigma_r^2$ respectively in Appendix C. Note that these parameters can be measured from engineering points of view, for example, by using the manufacturing datasheet of each hardware component of RF frontends in the real system [24].

### B. Downlink Transmission With Imperfect Channel Estimation

In TDD massive MIMO systems, UTs first transmit the orthogonal UL pilots to BS, which enables BS to estimate the UL channel. In this paper, we model the channel estimation error as the additive independent random error term [10], [12]. By taking the effect of $\mathbf{H}_{br}$ into consideration, the estimate $\hat{\mathbf{H}}_u$ of the actual uplink channel response $\mathbf{H}_u$ can be given by

$$\hat{\mathbf{H}}_u = \sqrt{1-\tau^2}\mathbf{H}_{br}\mathbf{H} + \tau\mathbf{V}, \quad (11)$$

where two $M \times K$ matrices $\mathbf{H}$ and $\mathbf{V}$ represent the propagation channel and the channel estimation error, respectively. We assume the entries of both $\mathbf{H}$ and $\mathbf{V}$ are independent identically distributed (i.i.d.) complex Gaussian random variables with zero mean and unit variance. In addition, the estimation variance parameter $\tau \in [0, 1]$ is applied to reflect the accuracy of the channel estimation, e.g., $\tau = 0$ represents the perfect estimation, whereas $\tau = 1$ corresponds to the case that the channel estimate is completely uncorrelated with the actual channel response.

The UL channel estimate $\hat{\mathbf{H}}_u$ is then exploited in the DL transmission for precoding. Specifically, by considering the channel reciprocity within the channel coherence period, the BS predicts the DL channel as

$$\hat{\mathbf{H}}_d = \hat{\mathbf{H}}_u^T = \sqrt{1-\tau^2}\mathbf{H}^T\mathbf{H}_{br} + \tau\mathbf{V}^T. \quad (12)$$

While the UL and DL propagation channels are reciprocal, the Tx and Rx frontends are not, due to the reciprocity error. By taking the effect of $\mathbf{H}_{bt}$ into the consideration, the actual DL channel $\mathbf{H}_d$ can be denoted as

$$\mathbf{H}_d = \mathbf{H}^T \mathbf{H}_{bt}. \tag{13}$$

Then, the BS performs the linear precoding for the DL transmission based on the DL channel estimate $\hat{\mathbf{H}}_d$ instead of the actual channel $\mathbf{H}_d$, and the received signal $\mathbf{y}$ for the $K$ UTs is given by

$$\mathbf{y} = \sqrt{\rho_d}\lambda \mathbf{H}_d \mathbf{W}\mathbf{s} + \mathbf{n} = \sqrt{\rho_d}\lambda \mathbf{H}^T \mathbf{H}_{bt} \mathbf{W}\mathbf{s} + \mathbf{n}, \tag{14}$$

where $\mathbf{W}$ represents the linear precoding matrix, which is a function of the DL channel estimate $\hat{\mathbf{H}}_d$ instead of the actual DL channel $\mathbf{H}_d$. The parameter $\rho_d$ denotes the average transmit power at the BS, and note that the power is equally allocated to each UT in this work. The vector $\mathbf{s}$ denotes the symbols to be transmitted to $K$ UTs. We assume that the symbols for different users are independent, and constrained with the normalised symbol power per user. To offset the impact of the precoding matrix on the transmit power, it is multiplied by a normalisation parameter $\lambda$, such that $\mathbb{E}\left\{\text{tr}\left(\lambda^2 \mathbf{W}\mathbf{W}^H\right)\right\} = 1$. This ensures that the transmit power after precoding remains equal to the transmit power budget that $\mathbb{E}\left\{\|\sqrt{\rho_d}\lambda \mathbf{W}\mathbf{s}\|^2\right\} = \rho_d$. In addition, $\mathbf{n}$ is the additive white Gaussian noise (AWGN) vector, whose $k$-th element is complex Gaussian distributed with zero mean and covariance $\sigma_k^2$, i.e., $n_k \sim \mathcal{CN}(0, \sigma_k^2)$. We assume that $\sigma_k^2 = 1$, $k = 1, 2, \cdots, K$. Therefore, $\rho_d$ can also be treated as the DL transmit signal-to-noise ratio (SNR).

By comparing the channel estimate $\hat{\mathbf{H}}_d$ for the precoding matrix in (12) with the actual DL channel $\mathbf{H}_d$ in (13), we have

$$\mathbf{H}_d = \frac{1}{\sqrt{1-\tau^2}}(\hat{\mathbf{H}}_d - \tau \mathbf{V}^T) \cdot \underbrace{\mathbf{H}_{br}^{-1}\mathbf{H}_{bt}}_{\text{reciprocity errors}}, \tag{15}$$

where the term $\mathbf{H}_{br}^{-1}\mathbf{H}_{bt}$ stands for the reciprocity errors, and is equivalent to $\mathbf{E}$ defined in (5) (also corresponds to the error model $\mathbf{E}_b$ in [20]). The expression (15) reveals that the channel reciprocity error is multiplicative, in the sense that the corresponding error term $\mathbf{H}_{br}^{-1}\mathbf{H}_{bt}$ is multiplied with the channel estimate $\hat{\mathbf{H}}_d$ and the estimation error $\mathbf{V}$. Based on the discussion followed by (5), $\mathbf{H}_d$ and $\hat{\mathbf{H}}_d$ can have one scale difference in the case that $\mathbf{H}_{br}^{-1}\mathbf{H}_{bt} = c\mathbf{I}_M$, thus no reciprocity error caused in this case. On the contrary, in the presence of the mismatch between $\mathbf{H}_{br}$ and $\mathbf{H}_{bt}$, the channel reciprocity error can be introduced into the system. From (15), it is also indicated that the integration between the multiplicative reciprocity error and the additive estimation error brings a compound effect on the precoding matrix calculation. We shall analyse this effect in the following Section III.

In order to investigate the effect of reciprocity errors on the performance of the linearly precoded system in terms of the output SINR for a given $k$-th UT, let $M \times 1$ vectors $\mathbf{h}_k$ and $\mathbf{v}_k$ be the $k$-th column of the channel matrix $\mathbf{H}$ and the estimation error matrix $\mathbf{V}$ respectively, as well as $\mathbf{w}_k$ and $s_k$ represent the precoding vector and the transmit symbol for the $k$-th UT, while $\mathbf{w}_i$ and $s_i$, $i \neq k$ for other UTs, respectively. Specifying the received signal $\mathbf{y}$ by substituting $\mathbf{h}_k$, $\mathbf{w}_k$ and $\mathbf{w}_i$ into (14), we rewrite the received signal for the $k$-th UT as

$$y_k = \underbrace{\sqrt{\rho_d}\lambda \mathbf{h}_k^T \mathbf{H}_{bt}\mathbf{w}_k s_k}_{\text{Desired Signal}} + \underbrace{\sqrt{\rho_d}\lambda \sum_{i=1,i\neq k}^{K} \mathbf{h}_k^T \mathbf{H}_{bt}\mathbf{w}_i s_i}_{\text{Inter-user Interference}} + \underbrace{n_k}_{\text{Noise}}. \tag{16}$$

The first term of the received signal $y_k$ in (16) is related to the desired signal for the $k$-th UT, and the second term represents the inter-user interference among other $K-1$ UTs. Then, the desired signal power $P_s$ and the interference power $P_I$ can be expressed as

$$P_s = |\sqrt{\rho_d}\lambda \mathbf{h}_k^T \mathbf{H}_{bt}\mathbf{w}_k s_k|^2, \tag{17}$$

$$P_I = \left|\sum_{i=1,i\neq k}^{K} \sqrt{\rho_d}\lambda \mathbf{h}_k^T \mathbf{H}_{bt}\mathbf{w}_i s_i\right|^2, \tag{18}$$

respectively. Considering (17), (18) and the third term in (16) which is the AWGN, the output SINR for the $k$-th UT in the presence of the channel reciprocity error can be given as in [25]

$$\text{SINR}_k = \mathbb{E}\left\{\frac{P_s}{P_I + \sigma_k^2}\right\} \approx \mathbb{E}\{P_s\} \mathbb{E}\left\{\frac{1}{P_I + \sigma_k^2}\right\}, \tag{19}$$

thus we can approximate the output SINR by calculating $\mathbb{E}\{P_s\}$ and $\mathbb{E}\{1/(P_I + \sigma_k^2)\}$ separately. In order to derive the term $\mathbb{E}\{1/(P_I + \sigma_k^2)\}$ and pursue the calculation of (19), we provide one generalised conclusion as in the following proposition.

*Proposition 2:* Let a random variable $X_1 \in \mathbb{C}$ and $X_1 \neq 0$, $\exists \left(\mathbb{E}\{X_1\}, \text{var}(X_1), \mathbb{E}\left\{\frac{1}{X_1}\right\}\right) \in \mathbb{C}$, and $\mathbb{E}\{X_1\} \neq 0$, then

$$\mathbb{E}\left\{\frac{1}{X_1}\right\} = \frac{1}{\mathbb{E}\{X_1\}} + O\left(\frac{\text{var}(X_1)}{\mathbb{E}\{X_1\}^3}\right). \tag{20}$$

*Proof:* Consider the Taylor series of $\mathbb{E}\left\{\frac{1}{X_1}\right\}$, we have

$$\mathbb{E}\left\{\frac{1}{X_1}\right\} = \mathbb{E}\left\{\frac{1}{\mathbb{E}\{X_1\}} - \frac{1}{\mathbb{E}\{X_1\}^2}(X_1 - \mathbb{E}\{X_1\}) + \frac{1}{\mathbb{E}\{X_1\}^3}(X_1 - \mathbb{E}\{X_1\})^2 - \cdots\right\}. \tag{21}$$

Then one can easily arrive at (20). □

From Proposition 2, it is expected that the approximation in (19) can be more precise than the widely-used approximate SINR expressions in the literatures, e.g., [18, eq. (6)] and [26, eq. (6)], which are based on $\text{SINR}_k \approx \mathbb{E}\{P_s\}/\mathbb{E}\{P_I + \sigma_k^2\}$ that is not accurate when the value of $(\text{var}(X_1)/\mathbb{E}\{X_1\}^3)$ is not negligible. We will verify the accuracy of (19) in the analytical results in the following section.

## III. SINR for Maximum-Ratio Transmission and Zero-Forcing Precoding Schemes

In this section, we formulate and discuss the effect of the reciprocity error on the performance of MRT and ZF precoding schemes, in terms of the output SINR, by considering the reciprocity error model with the truncated Gaussian distribution.

### A. Maximum-Radio Transmission

Recall (12) and (14), for MRT, the precoding matrix $\mathbf{W}$ can be given by

$$\mathbf{W}_{\text{mrt}} = \hat{\mathbf{H}}_d^H = \sqrt{1-\tau^2}\mathbf{H}_{br}^*\mathbf{H}^* + \tau\mathbf{V}^*. \tag{22}$$

Let $\lambda_{\text{mrt}}$ represent the normalisation parameter of the MRT precoding scheme to meet the power constraint, which can be calculated as

$$\lambda_{\text{mrt}} = \sqrt{\frac{1}{\mathbb{E}\{\text{tr}(\mathbf{W}_{\text{mrt}}\mathbf{W}_{\text{mrt}}^H)\}}} = \sqrt{\frac{1}{MK((1-\tau^2)A_r + \tau^2)}}. \tag{23}$$

The proof of (23) is briefed in Appendix D. For the sake of simplicity, we define the amplitude-error-related factors $A_r$ in (23) and $A_t$ as

$$A_r \triangleq \alpha_r^2 + \sigma_r^2, \quad A_t \triangleq \alpha_t^2 + \sigma_t^2, \tag{24}$$

and we assume the small deviation of the amplitude errors [15], i.e., $A_t$, $A_r \approx 1$. In addition, let $A_I$ be the aggregated reciprocity error factor, which can be given by

$$A_I \triangleq \frac{\alpha_t^2 \alpha_r^2}{(\alpha_t^2 + \sigma_t^2)(\alpha_r^2 + \sigma_r^2)}|g_t|^2|g_r|^2, \tag{25}$$

where $\alpha_t^2$, $\alpha_r^2$, $\sigma_r^2$ and $\sigma_t^2$ as well as $g_t$ and $g_r$ are given following Proposition 1, and detailed in Appendix C. Based on the values of $\alpha_t$, $\alpha_r$, $\sigma_t^2$, $\sigma_r^2$, $g_t$ and $g_r$, we have $0 < A_I \le 1$. More specifically, when the level of the channel reciprocity errors decreases in the system, we have $\alpha_t, \alpha_r, g_t, g_r \to 1$ and $\sigma_r^2, \sigma_t^2 \to 0$, thus $A_I \to 1$. And the perfect channel reciprocity corresponds to $A_I = 1$. In contrast, when the level of the reciprocity errors increases, we have $A_I \to 0$.

By using (17), (23), (24) and (25), the expected value of the desired signal power $P_{s,\text{mrt}}$ can be given as

$$\mathbb{E}\{P_{s,\text{mrt}}\} = \mathbb{E}\{|\sqrt{\rho_d}\lambda_{\text{mrt}}\mathbf{h}_k^T\mathbf{H}_{bt}\mathbf{w}_{k,\text{mrt}}s_k|^2\}$$
$$= \frac{\rho_d A_t}{K}\left(\frac{(1-\tau^2)A_r((M-1)A_I + 2) + \tau^2}{(1-\tau^2)A_r + \tau^2}\right). \tag{26}$$

Similarly, the expectation of interference power $P_{I,\text{mrt}}$ can be computed based on (18) and (23) as

$$\mathbb{E}\{P_{I,\text{mrt}}\} = \mathbb{E}\left\{\left|\sum_{i=1,i\neq k}^{K}\sqrt{\rho_d}\lambda_{\text{mrt}}\mathbf{h}_k^T\mathbf{H}_{bt}\mathbf{w}_{i,\text{mrt}}s_i\right|^2\right\}$$
$$= \rho_d\frac{K-1}{K}A_t. \tag{27}$$

The proof of (26) and (27) can be found in Appendix D.

Based on (23), (26), (27) and (19) with Proposition 2, the analytical expression of the output SINR for the $k$-th UT with MRT precoder can be obtained as in the following theorem.

*Theorem 1:* Consider a massive MIMO system with $K$ UTs and $M$ BS antennas, and the propagation channel follows the i.i.d. standard complex Gaussian random distribution. The channel estimation error is modelled as the additive independent Gaussian variables. The MRT precoding scheme is used at the BS. The channel reciprocity error is brought by the mismatch between the RF frontends matrices of the Tx-front $\mathbf{H}_{bt}$ and the Rx-front $\mathbf{H}_{br}$, where both amplitude and phase components of the diagonal entries are followed the truncated Gaussian random distribution. Then the closed-form expression of the output SINR for the $k$-th UT is given by

$$\text{SINR}_{k,\text{mrt}} \approx \mathbb{E}\{P_{s,\text{mrt}}\}\mathbb{E}\left\{\frac{1}{P_{I,\text{mrt}} + \sigma_k^2}\right\} \tag{28}$$

$$= \rho_d A_t\left(\frac{(1-\tau^2)A_r((M-1)A_I+2)+\tau^2}{(1-\tau^2)A_r+\tau^2}\right)$$
$$\times\left(\frac{K^2 + \rho_d K(K-1)(\rho_d A_t^2 + 2A_t)}{(\rho_d(K-1)A_t + K)^3}\right), \tag{29}$$

where $A_I$ is given by (25), and $A_t$ as well as $A_r$ are defined in (24).

*Proof:* Let $X_{1,\text{mrt}} \triangleq P_{I,\text{mrt}} + \sigma_k^2$, and the term $\mathbb{E}\{1/(P_{I,\text{mrt}} + \sigma_k^2)\}$ can be calculated based on Proposition 2. Specifically, in our case, we have

$$\mathbb{E}\{X_{1,\text{mrt}}\} = \mathbb{E}\{P_{I,\text{mrt}}\} + \mathbb{E}\{\sigma_k^2\} = \rho_d\frac{K-1}{K}A_t + 1, \tag{30}$$

and

$$\text{var}(X_{1,\text{mrt}})$$
$$= \text{var}(P_{I,\text{mrt}}) + \text{var}(\sigma_k^2) \tag{31}$$
$$= \text{var}\left(\left|\sum_{i=1,i\neq k}^{K}\sqrt{\rho_d}\lambda_{\text{mrt}}\mathbf{h}_k^T\mathbf{H}_{bt}(\sqrt{1-\tau^2}\mathbf{H}_{br}^*\mathbf{h}_i^* + \tau\mathbf{v}_i^*)s_i\right|^2\right) \tag{32}$$
$$= \rho_d^2\lambda_{\text{mrt}}^4(K-1)\left(2\mathbb{E}\{\tilde{X}_1\}^2\text{var}(\tilde{X}_1) + \text{var}(\tilde{X}_1)^2\right), \tag{33}$$

where $\tilde{X}_1 \triangleq \mathbf{h}_k^T\mathbf{H}_{bt}(\sqrt{1-\tau^2}\mathbf{H}_{br}^*\mathbf{h}_i^* + \tau\mathbf{v}_i^*)$, and $\mathbb{E}\{\tilde{X}_1\}$ and $\text{var}(\tilde{X}_1)$ are given as

$$\mathbb{E}\{\tilde{X}_1\} = \sqrt{1-\tau^2}\mathbb{E}\{\mathbf{h}_k^T\mathbf{H}_{bt}\mathbf{H}_{br}^*\mathbf{h}_i^*\} + \tau\mathbb{E}\{\mathbf{h}_k^T\mathbf{H}_{bt}\mathbf{v}_i^*\}$$
$$= 0, \tag{34}$$
$$\text{var}(\tilde{X}_1) = \mathbb{E}\{|\mathbf{h}_k^T\mathbf{H}_{bt}(\sqrt{1-\tau^2}\mathbf{H}_{br}^*\mathbf{h}_i^* + \tau\mathbf{v}_i^*)|^2\}$$
$$= MA_t\left((1-\tau^2)A_r + \tau^2\right). \tag{35}$$

Hence, substituting (34) and (35) into (33), the complete result of (33) is obtained. Next, applying (30) and the completed (33) to (20) in Proposition 2 yields the term

$$\frac{\text{var}(X_{1,\text{mrt}})}{\mathbb{E}\{X_{1,\text{mrt}}\}^3} = \frac{\rho_d^2(K-1)KA_t^2}{(\rho_d(K-1)A_t + K)^3}. \tag{36}$$

By using (30) and (36), $\mathbb{E}\{1/(P_{I,\text{mrt}} + \sigma_k^2)\}$ is obtained, which can then be substituted into (28) together with (26). We now arrive at (29). □

MI *et al.*: MASSIVE MIMO PERFORMANCE WITH IMPERFECT CHANNEL RECIPROCITY AND CHANNEL ESTIMATION ERROR 3739From (36) in Theorem 1, it is expected that the value of $\left(\mathrm{var}(X_{1,\mathrm{mrt}})/\mathbb{E}\{X_{1,\mathrm{mrt}}\}^3\right)$ can be negligible in the case with the large number of UTs or in the high SNR regime, and based on (20) in Proposition 2, the result (29) can be simplified to [26, eq. (13)] in the absence of reciprocity error and estimation error. However, in the low SNR regime or $K$ is small, the approximation $\mathrm{SINR}_k \approx \mathbb{E}\{P_s\}/\mathbb{E}\{P_I + \sigma_k^2\}$ becomes less accurate due to the significant value of $\left(\mathrm{var}(X_{1,\mathrm{mrt}})/\mathbb{E}\{X_{1,\mathrm{mrt}}\}^3\right)$. Hence, we use the approximate SINR expression in (19) in this paper, for more generic cases of TDD massive MIMO systems. In addition, more detailed discussions of (26), (27) and (29) will be provided at the end of this section.

*B. Zero-Forcing*

Similar to MRT, the precoding matrix for the ZF precoded system can be written as

$$\mathbf{W}_{\mathrm{zf}} = \hat{\mathbf{H}}_d^H \left(\hat{\mathbf{H}}_d \hat{\mathbf{H}}_d^H\right)^{-1}, \quad (37)$$

where $\hat{\mathbf{H}}_d$ is given in (12). The corresponding normalisation parameter can be given as

$$\lambda_{\mathrm{zf}} = \sqrt{\frac{1}{\mathbb{E}\left\{\mathrm{tr}\left(\mathbf{W}_{\mathrm{zf}}\mathbf{W}_{\mathrm{zf}}^H\right)\right\}}} \approx \sqrt{\frac{M-K}{K}\left((1-\tau^2)A_r + \tau^2\right)}, \quad (38)$$

and be used to satisfy the power constraint. The proof of (38) is given in Appendix E. Then two propositions can be provided to present the performance of the desired signal power and the interference power as follows.

*Proposition 3:* Let the similar assumptions be held as in Theorem 1, and ZF precoding scheme be implemented in the system. For a given UT $k$, the expectation of the signal power in the presence of the reciprocity error can be expressed as

$$\mathbb{E}\{P_{s,\mathrm{zf}}\} = \mathbb{E}\left\{|\sqrt{\rho_d}\lambda_{\mathrm{zf}}\mathbf{h}_k^T\mathbf{H}_{bt}\mathbf{w}_{k,\mathrm{zf}}s_k|^2\right\}$$
$$\approx \rho_d \frac{M-K}{K}B_I, \quad (39)$$

where the error parameter $B_I$ can be defined by

$$B_I \triangleq \frac{(1-\tau^2)A_I A_t A_r}{(1-\tau^2)A_r + \tau^2}. \quad (40)$$

*Proof:* See Appendix E in detail. □

*Proposition 4:* Let the same conditions be assumed as in Proposition 3. For a given UT $k$, the expectation of the inter-user-interference power can be given as

$$\mathbb{E}\{P_{I,\mathrm{zf}}\} = \mathbb{E}\left\{\left|\sum_{i=1,i\neq k}^{K}\sqrt{\rho_d}\lambda_{\mathrm{zf}}\mathbf{h}_k^T\mathbf{H}_{bt}\mathbf{w}_{i,\mathrm{zf}}s_i\right|^2\right\}$$
$$\approx \rho_d \frac{K-1}{K}(A_t - B_I). \quad (41)$$

*Proof:* See Appendix E. □

Combine the results in Proposition 3 and 4, we can derive the theoretical expression of the output SINR for the $k$-th UT in the ZF precoded system as following.

*Theorem 2:* In a ZF precoded system, by assuming that the same conditions are held as in Theorem 1, the output SINR for the $k$-th UT under the effect of the reciprocity error, can be formulated as

$$\mathrm{SINR}_{k,\mathrm{zf}}$$
$$\approx \mathbb{E}\{P_{s,\mathrm{zf}}\}\mathbb{E}\left\{\frac{1}{P_{I,\mathrm{zf}}+\sigma_k^2}\right\} \quad (42)$$
$$\approx \rho_d(M-K)B_I$$
$$\times \left(\frac{K^2 + \rho_d K(K-1)(A_t - B_I)(\rho_d(A_t - B_I)+2)}{(\rho_d(K-1)(A_t - B_I)+K)^3}\right), \quad (43)$$

where $B_I$ is defined in (40), and $A_t$ can be found in (24).

*Proof:* Consider the same method as shown in the proof of Theorem 1 based on Proposition 2, let $X_{1,\mathrm{zf}} = P_{I,\mathrm{zf}} + \sigma_k^2$ and we have

$$\mathbb{E}\{X_{1,\mathrm{zf}}\} = \mathbb{E}\{P_{I,\mathrm{zf}}\} + \mathbb{E}\{\sigma_k^2\} \approx \rho_d \frac{K-1}{K}(A_t - B_I) + 1, \quad (44)$$

based on (41). Following the discussions of $\mathbb{E}\{P_{I,\mathrm{zf}}\}$ in Appendix E, we have

$$\mathrm{var}(X_{1,\mathrm{zf}}) = \mathrm{var}(P_{I,\mathrm{zf}}) + \mathrm{var}(\sigma_k^2) \approx \rho_d^2 \frac{K-1}{K^2}(A_t - B_I)^2. \quad (45)$$

Substituting (44) and (45) into (20), $\mathbb{E}\{1/(P_{I,\mathrm{zf}}+\sigma_k^2)\}$ can be obtained. Together with (39), we have (43). □

Similar to the discussion following Theorem 1, the expression (43) can be simplified into the corresponding result in [10, eq. (44)] by considering the perfect channel reciprocity and the large number of UTs. Furthermore, our expression in (43) can be applied in more generic cases, e.g., $K$ is small.

To this end, the analytical expressions of the output SINR in the MRT and ZF precoded systems are provided in (29) and (43) respectively. Note that the deduction from the results in the Theorem 1 and Theorem 2 to specialised cases such as the general Gaussian distributed errors can be straightforward, simply by setting the truncated ranges to infinity. We shall provide the analysis and comparison of these expressions in the following discussions.

*C. Discussions*

We first consider the impact of the reciprocity error on the desired signal power and interference power separately. For the MRT precoded system, it can be observed from (26) and (27) that both Tx/Rx-front phase errors degrade the desired signal power, but neither of them contributes to the interference power since non-coherent adding of the precoder and the channel for the interference. Move on to the amplitude errors, only the Tx-front error exists in (27), and amplifies the interference power, which is unlike the impact on the signal power, where both Tx/Rx front amplitude errors are present. Recall (39) and (41) for the ZF precoded system, apparently, both the desired signal power and the inter-user interference power are affected by the amplitude and phase reciprocity errors at both Tx/Rx frontends.

We then take the channel estimation error into account. Based on (29) and (43), an intuitive conclusion can be drawn that the increase of the estimation error results in the performance degradation of the output SINR, for both MRT and ZF. Furthermore, it is expected that the effect of the estimation error may be amplified by the reciprocity error, in the sense that the estimation error is multiplied with the reciprocity error as shown in (15).

Note that the focus of this paper is to investigate the effect of imperfect channel reciprocity on the performance of MRT and ZF precoding schemes. We remove the channel estimation error from (29) in Theorem 1 with (43) in Theorem 2, i.e., let $\tau$ be zero, and obtain

$$\tilde{\text{SINR}}_{k,\text{mrt}} \approx \rho_d \left(((M-1)A_I A_t + 2A_t)\right) \times \left(\frac{K^2 + \rho_d K(K-1)(\rho_d A_t^2 + 2A_t)}{(\rho_d(K-1)A_t + K)^3}\right), \quad (46)$$

and

$$\tilde{\text{SINR}}_{k,\text{zf}} \approx \rho_d (M-K) A_I A_t \times \left(\frac{K^2 + \rho_d K(K-1)A_t(1-A_I)(\rho_d A_t(1-A_I) + 2)}{(\rho_d(K-1)A_t(1-A_I) + K)^3}\right), \quad (47)$$

where $\tilde{\text{SINR}}_{k,\text{mrt}}$ and $\tilde{\text{SINR}}_{k,\text{zf}}$ represent the output SINR under the effect of the reciprocity error only. Comparing (46) with (47), first, we observe that the effects of the Tx and Rx front amplitude errors are not equivalent for both MRT and ZF, thus it is meaningful to model $\mathbf{H}_{br}$ and $\mathbf{H}_{bt}$ separately. Second, it can be claimed that the ZF precoding scheme is likely to be more sensitive to the phase errors compared to MRT. For example, Due to the phase error involving in the ZF precoded system, the power of the desired signal decreases and the power of the interference increases, whereas no effect of the phase error on the interference power when MRT is implemented. Hence, more impact of the phase errors on the ZF precoder can be expected than that on the MRT precoder.

## IV. ASYMPTOTIC SINR ANALYSIS

In this section, we simplify the closed-form expressions in Theorem 1 and Theorem 2, by considering the case when $M$ goes to infinity, which leads to several implications for the massive MIMO systems.

### A. Without Channel Estimation Error

We first focus on the expressions of $\tilde{\text{SINR}}_{k,\text{mrt}}$ and $\tilde{\text{SINR}}_{k,\text{zf}}$, and analyse the effect of the reciprocity error on the MRT and ZF precoded systems without considering the channel estimation error.

*1) Maximum Ratio Transmission:* Recall (46), two multiplicative terms are corresponded to the desired signal power and interference power. When $K \gg 1$, the second term becomes $\mathbb{E}\left\{1/(P_{I,\text{mrt}} + \sigma_k^2)\right\} \approx 1/(\rho_d A_t + 1)$, thus $\tilde{\text{SINR}}_{k,\text{mrt}}$ can be approximated by

$$\tilde{\text{SINR}}_{k,\text{mrt}} \xrightarrow{K \gg 1} \frac{\rho_d((M-1)A_I + 2)}{K(\rho_d + A_t^{-1})}. \quad (48)$$

In the high region of transmit SNR, by assuming $M \to \infty$ and $A_t \approx 1$ as mentioned in (24), the asymptotic expression of (46) can be given as

$$\lim_{\substack{M \to \infty, \\ K \gg 1}} \tilde{\text{SINR}}_{k,\text{mrt}} = \frac{M}{K} A_I, \quad (49)$$

where $A_I$ can be found in (25). As discussed in the paragraph following (25), we have $A_I = 1$ in the case with perfect channel reciprocity, whereas $A_I \to 0$ when the level of reciprocity errors increases.

From (49), several conclusions can be given for MRT. First, the asymptotic expression in (49) can be simplified to the result in [9, Table 1] in the case with the perfect channel reciprocity and high transmit SNR, and the output SINR of MRT is upper-bounded by the ratio $M/K$ due to the inter-user interference. Second, when the significant reciprocity error is introduced into the system, we have $A_I \to 0$, and consequently, the larger number of $M$ or increasing ratio of $M/K$ may not lead to the better system performance, due to the error ceiling limited by the reciprocity error, which corresponds the multiplicative term (i.e., $A_I$) that in (49).

*2) Zero Forcing Precoding:* Similar to (49), we update the analytical results of the output SINR for ZF, asymptotically with $M \to \infty$ and $K \gg 1$. Recall (47), we have

$$\lim_{\substack{M \to \infty, \\ K \gg 1}} \tilde{\text{SINR}}_{k,\text{zf}} = \frac{\rho_d(M-K) A_I A_t}{K(\rho_d A_t(1-A_I) + 1)}. \quad (50)$$

In the case with the perfect channel reciprocity, we have $A_I = 1$, then (50) can be transformed to the result in [9, Table 1]. Since $(1 - A_I) = 0$ in this case, it is unlikely to directly simplify (50) to the noise-free case (as (49) of MRT) even in the high region of transmit SNR. When the level of the reciprocity error increases, we have $(1 - A_I) > 0$. Consider a case with $\rho_d(1 - A_I) \gg 1$, which may be achieved with the high region of transmit SNR and the nontrivial value of $(1 - A_I)$, the denominator in (50) is dominated by $K\rho_d A_t(1-A_I)$. Assuming the large ratio of $M/K$, (50) can be further simplified as

$$\lim_{\substack{M \to \infty, M \gg K \gg 1, \\ \rho_d(1-A_I) \gg 1}} \tilde{\text{SINR}}_{k,\text{zf}} = \frac{M}{K} \left(\frac{1}{A_I^{-1} - 1}\right), \quad (51)$$

with $A_I^{-1} > 1$ in this case. From (50), again, we can conclude that the ZF precoded system performance can be hindered due to the impact of both amplitude and phase reciprocity errors, even with the infinite number of BS antennas. Also, when the higher level of the reciprocity error is introduced, the variation of the output SINR can be independent of the transmit SNR, and the error ceiling, which corresponds to the reciprocity-error-related multiplicative component $(1/(A_I^{-1} - 1))$, can be observed in (51).

*3) Comparison:* From (49) and (51), we observe that the channel reciprocity errors causes the random multiplicative distortions. One aspect of the error effects is the error ceilings, e.g., $A_I$ in (49) for MRT and $\left(1/\left(A_I^{-1} - 1\right)\right)$ in (51) for ZF. Besides the previous discussions followed by (49) and (51), several implications can be provided by comparing the performance of MRT and ZF. Consider the same assumption for (51), in the high region of the transmit SNR, we have

$$\tilde{C}_I \triangleq \lim_{\substack{M\to\infty, M\gg K\gg 1 \\ \rho_d(1-A_I)\gg 1}} \frac{\text{SI}\tilde{\text{N}}\text{R}_{k,\text{zf}}}{\text{SI}\tilde{\text{N}}\text{R}_{k,\text{mrt}}} = \frac{1}{1-A_I} > 1, \quad (52)$$

where the term $\tilde{C}_I$ denotes the ratio of the asymptotic SINR expressions of ZF and MRT. Under the conditions of (52), it can be concluded that the performance preponderance of using ZF over MRT is only conditioned on the level of reciprocity errors. In the case that $A_I \to 1$, the lower level of the reciprocity error is introduced into the systems, and ZF outperforms MRT in terms of the output SINR. On contrary, when $A_I \to 0$, corresponding to the significantly high level of the reciprocity errors, the ZF precoded system is more affected by the channel reciprocity errors than the MRT system, and consequently, the performance degradation of both systems results in the almost identical output SINR, which can be represented by

$$\tilde{C}_I \xrightarrow{A_I \to 0} 1. \quad (53)$$

This leads to a useful guidance of precoding schemes selection for the massive MIMO systems in the presence of channel reciprocity errors in practice.

### B. Imperfect Channel Estimation

We extend the prior analysis in (49) and (51) by considering the channel estimation error. Recall (29) in Theorem 1 and (43) in Theorem 2, and consider the same conditions for (49) and (50), we obtain the asymptotic expressions as

$$\lim_{\substack{M\to\infty, \\ K\gg 1}} \text{SINR}_{k,\text{mrt}} = \frac{M}{K}\left(\frac{\rho_d \tilde{B}_I}{\rho_d + A_t^{-1}}\right), \quad (54)$$

and

$$\lim_{\substack{M\to\infty, \\ K\gg 1}} \text{SINR}_{k,\text{zf}} = \frac{M-K}{K}\left(\frac{\rho_d \tilde{B}_I}{\rho_d(1-\tilde{B}_I) + A_t^{-1}}\right), \quad (55)$$

where $\tilde{B}_I \triangleq B_I A_t^{-1}$. Note that we have assumed that $A_t, A_r \approx 1$ in the discussion following (24), hence we approximate

$$\tilde{B}_I \approx (1-\tau^2)A_I. \quad (56)$$

From (56), it is expected that $\tilde{B}_I \to 0$ when $A_I \to 0$, irrespective of the existence of the estimation error. Then, similar to (52), we can define

$$C_I \triangleq \lim_{\substack{M\to\infty, \\ M\gg K\gg 1}} \frac{\text{SINR}_{k,\text{zf}}}{\text{SINR}_{k,\text{mrt}}} = \frac{1}{1-\tilde{B}_I} \geq 1, \quad (57)$$

where $C_I$ is the generalised expression of $\tilde{C}_I$, by taking the imperfect channel estimation into the consideration. Note that in (57), the case with $C_I = 1$ corresponds to that the channel estimate and actual channel are uncorrelated, i.e., $\tau = 1$. In addition, we can conclude that

$$C_I \xrightarrow{A_I \to 0} 1. \quad (58)$$

Therefore, the conclusion following (52) still holds when the channel estimation error is introduced into the system.

## V. SIMULATION RESULTS

In this section, we present simulation results to compare the performance of ZF and MRT precoders in massive MIMO systems with reciprocity errors, and validate the analytical expressions of the output SINR in Section III and asymptotic results of Section IV. Unless specified otherwise, the number of BS antennas $M = 500$, the number of single-antenna UTs $K = 20$, and the transmit SNR, $\rho_d = 10$ dB (note that equal power allocation is considered for $K$ UTs). We model the random variables $A_{br,i}$, $A_{bt,i}$, $\varphi_{br,i}$ and $\varphi_{bt,i}$ as independent truncated Gaussian distribution. In order to clarify the combinations of the parameters for each random variable, e.g., the expected value $\alpha_{br,0}$, variance $\sigma_{br}^2$ and truncated ranges $[a_r, b_r]$ for $A_{br,i}$, we use quadruple notations, e.g., $(\alpha_{br,0}, \sigma_{br}^2, [a_r, b_r])$, and similar terms apply for $A_{bt,i}$, $\varphi_{br,i}$ and $\varphi_{bt,i}$. These parameters that related to amplitude and phase errors are measured in dB and in degrees (denoted by $(\cdot)°$), respectively.

### A. Channel Reciprocity Error Only

The focus of this paper is on the effect of the reciprocity errors on the system performance; hence, we first present the simulation results corresponding to the expressions (46) and (47).

*1) SINR Analysis for MRT and ZF:* To verify the theoretical results of $\text{SI}\tilde{\text{N}}\text{R}_{k,\text{mrt}}$ and $\text{SI}\tilde{\text{N}}\text{R}_{k,\text{zf}}$, we first consider a special case where only the amplitude mismatch error is present. Here, since the effects of $\varphi_{br,i}$ and $\varphi_{bt,i}$ are equivalent on (46) and (47), we introduce the constant phase error with $(\theta_{br,0}, \sigma_{\varphi_r}^2, [\theta_{r,1}, \theta_{r,2}]) = (\theta_{bt,0}, \sigma_{\varphi_t}^2, [\theta_{t,1}, \theta_{t,2}]) = (0°, 0.5, [-20°, 20°])$ following [20]. Let the amplitude error variances $\sigma_{br}^2 = \sigma_{bt}^2 = \sigma_A^2$ be the x-axis, the effect of the amplitude errors $A_{br,i}$ and $A_{bt,i}$ on MRT and ZF can be given in Fig. 2 and Fig. 3 respectively, where we consider the following scenarios:

Case 1: Considering certain parameters, e.g., changing the truncated range $[a_r, b_r]$ and the variance $\sigma_A^2$ in Fig. 2(a) and Fig. 3(a).

Case 2: Comparing the impacts of Tx and Rx frontends, e.g., for MRT, Fig. 2(b) vs Fig. 2(d); for ZF, Fig. 3(b) vs Fig. 3(d).

Case 3: Comparing the error impacts on ZF and MRT, e.g., Fig. 2(a) vs Fig. 3(a).

By considering the various amplitude error parameters (as Case 1) in Fig. 2 and Fig. 3, our analytical results exactly match the simulated results for both MRT and ZF. Additionally, considering the above scenarios, we observe the following:

OB1. For both ZF and MRT, the impact of the Tx front amplitude errors is different from that of the Rx front. For example, in Case 2, the results of Fig. 2(a) and Fig. 2(c) show a slight difference between the truncated ranges of amplitude errors $[a_r, b_r]$ and $[a_t, b_t]$, while

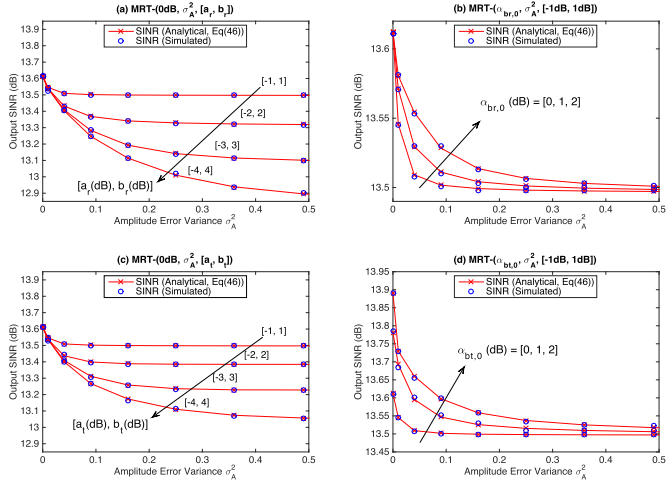

Fig. 2. Output SINR with MRT precoding in the presence of fixed phase errors and different combinations of amplitude errors.

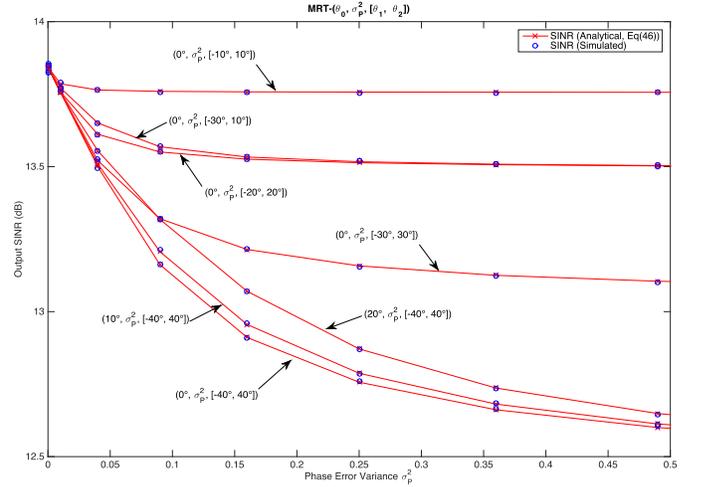

Fig. 4. Output SINR with MRT precoding in the presence of fixed amplitude errors and different combinations of phase errors.

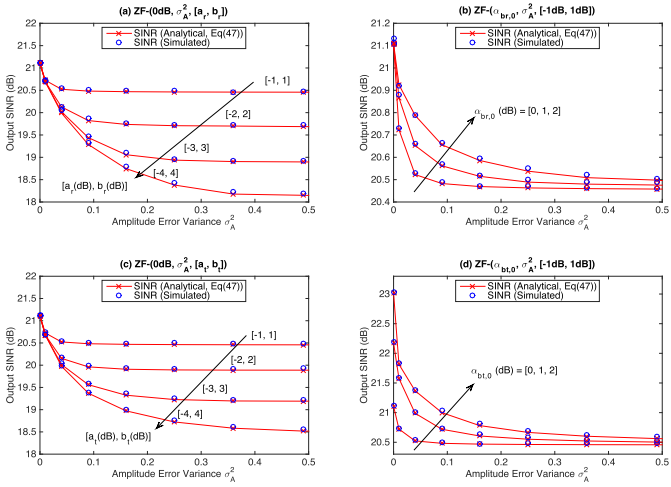

Fig. 3. Output SINR with ZF precoding in the presence of fixed phase errors and different combinations of amplitude errors.

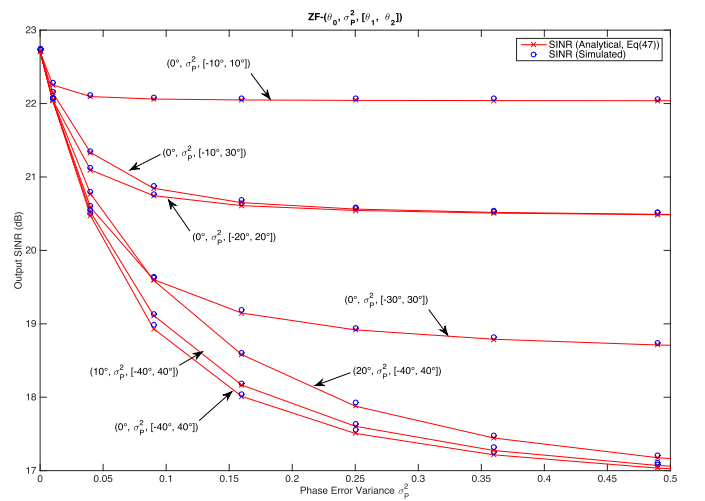

Fig. 5. Output SINR with ZF precoding in the presence of fixed amplitude errors and different combinations of phase errors.

Fig. 2(b) vs Fig. 2(d) demonstrate a greater impact from the expected value of Tx front amplitude errors $\alpha_{bt,0}$ than that from Rx front $\alpha_{br,0}$.

- OB2. It can be revealed from Case 3 that ZF is much more sensitive to the amplitude errors than MRT, as we discussed in Section IV. For example, comparing Fig. 3(a) and Fig. 2(a), with the same parameters, ZF experiences nearly 3 dB SINR loss compared to less than 1 dB loss in MRT.

Moving on to the phase reciprocity error, we fix the amplitude errors to $(\alpha_{br,0}, \sigma_{br}^2, [a_r, b_r]) = (\alpha_{bt,0}, \sigma_{bt}^2, [a_t, b_t]) = (0\ \text{dB}, 0.5, [-1\ \text{dB}, 1\ \text{dB}])$ as in [20]. As shown in (46) and (47), the phase errors $\varphi_{br,i}$ and $\varphi_{bt,i}$ have similar effect on SINR, hence, we assume $(\theta_{br,0}, \sigma_{\varphi_r}^2, [\theta_{r,1}, \theta_{r,2}]) = (\theta_{bt,0}, \sigma_{\varphi_t}^2, [\theta_{t,1}, \theta_{t,2}]) = (\theta_0, \sigma_P^2, [\theta_1, \theta_2])$ as shown in Fig. 4 and Fig. 5.

The perfect match between the simulation results and our analytical results can be observed from Fig. 4 and Fig. 5. We also draw the following observations:

- OB3. From Fig. 5, the phase errors can cause significant degradation of the ZF precoded system, e.g., with $(0°, 0.5, [-40°, 40°])$, almost 6 dB loss in terms of SINR, whereas the less severe SINR degradation (around 2 dB loss) can be seen from Fig. 4 for the MRT system.
- OB4. The main factors of the phase error are likely to be the error variance $\sigma_P^2$ and the relative truncated range, i.e., $(\theta_2 - \theta_1)$, rather than the expected values $\theta_0$ (see the closed curves between which the only difference is the increased expected values $0°$, $10°$ and $20°$ in Fig. 4, and similar in Fig. 5) and the absolute values of $\theta_1$ and $\theta_2$ (see the closed curves with truncated ranges $[-30°, 10°]$ and $[-20°, 20°]$ in Fig. 4, and with $[-10°, 30°]$ and $[-20°, 20°]$ in Fig. 5).

To summarise, it can be observed that the MRT precoded system is more tolerant to both amplitude and phase reciprocity errors compared with ZF, which is consistent with the theoretical analysis in Section III-C.

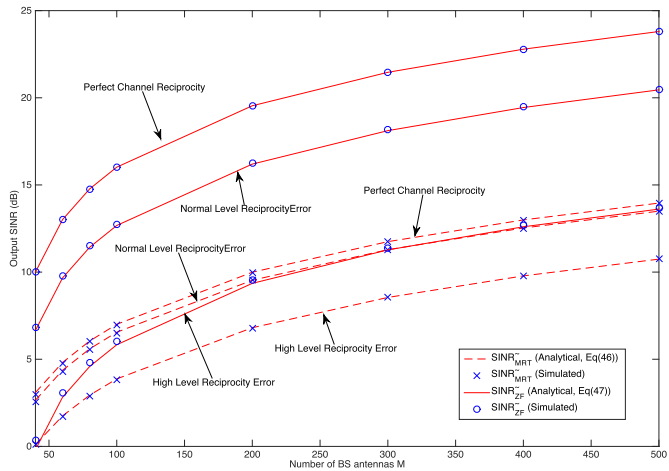

Fig. 6. Output SINR versus $M$ in the presence of different levels of channel reciprocity errors.

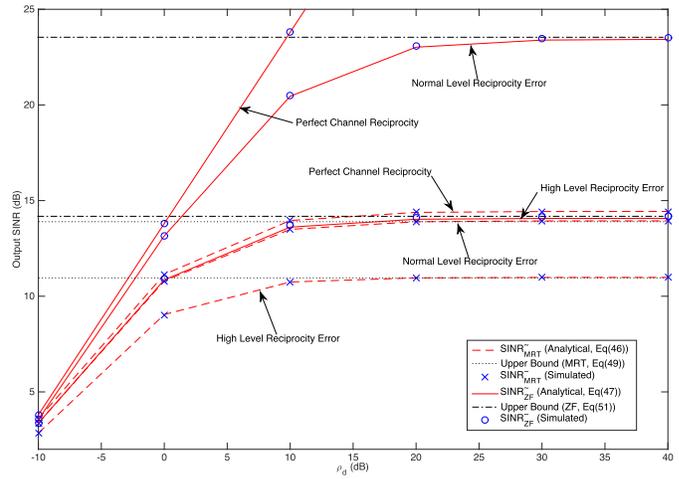

Fig. 7. Output SINR versus SNR in the presence of different levels of channel reciprocity errors.

*2) When M Goes to Infinity:* The theoretical results in Theorem 1 and 2, as well as (46) and (47) are conditioned on a large number of BS antennas $M$, which motivates us to investigate the case with the asymptotic limit, i.e., $M \to \infty$. Again, for the sake of easy comparison with the previous simulation results, let the same error parameters be considered for the transmit and receive sides. Also, we define the "Normal Level Reciprocity Error" with the amplitude errors $(0 \text{ dB}, 0.5, [-1 \text{ dB}, 1 \text{ dB}])$ and phase errors $(0°, 0.5, [-20°, 20°])$ as considered in [20], and "High Level Reciprocity Error" with $(0 \text{ dB}, 1, [-4 \text{ dB}, 4 \text{ dB}])$ and $(0°, 1, [-50°, 50°])$.

Fig. 6 demonstrates the performance of the output SINR for ZF and MRT with different values of $M$. It can be concluded that our theoretical results accurately reflect the system performance in all cases, even with the *not-so-large* values of $M$ comparing to $K$ (e.g., $M \leq 50$), which corresponds to the theory in [27]. Also, in general, ZF outperforms MRT, but again, it is much less tolerant to reciprocity errors. Specifically, with high-level errors, more than 10 dB SINR degradation is observed in the ZF precoded system, compared to the system with the ideal channel reciprocity.

Fig. 7 investigates the error ceiling effect that discussed in Section IV by increasing the transmit SNR $\rho_d$. We have $M = 500$, $K = 20$ to satisfy the conditions of the limit that $M \to \infty$ and $K \gg 1$. Without the channel reciprocity errors, the output SINR of ZF rises without an upper bound as growth of $\rho_d$, while that of MRT suffers from the inter-user interference in the high regime of $\rho_d$. The error ceiling obtained in Fig. 7 match the result in (49) for MRT, and the result in (51) for ZF. This, in turn, leads to the conclusion that in the high regime SNR (e.g., $\rho_d \geq 20$ dB), both ZF and MRT suffer from the impact of the reciprocity errors, which results in the degraded performance that is independent of the transmit SNR. In addition, we observe from Fig. 6 and 7 that MRT outperforms ZF in the low SNR regime or with the relatively small ratio of $M/K$.

*B. Imperfect Channel Estimation*

We then extend our investigations in Fig. 7 by taking the channel estimation error into considerations. The same conditions are applied as in Fig. 7, in addition with the estimation error parameter $\tau^2 = 0.1$. As shown in Fig. 8, the close match between the analytical and simulated results validates the output SINR expressions in (29) for MRT and (43) for ZF, as well as the error ceiling factors in (54) and (55). Furthermore, it reveals the significant impact of the reciprocity error on the estimation error. For example, in the case that $\rho_d = 10$dB, the estimation error (with $\tau^2 = 0.1$) causes slight performance degradation of the output SINR of the MRT precoder, around 0.5dB, which is then considerably increased to 4dB when the high-level reciprocity error introduced. The ZF precoded system with imperfect channel estimation suffers more from the reciprocity errors, such that more than 10 dB SINR loss can be experienced in the case with the high-level reciprocity error, compared with the degraded performance caused by the estimation error only. In addition, the results in Fig. 7 and 8 can be considered in selecting suitable modulation schemes for the practical massive MIMO system in the presence of different levels of the reciprocity error and the estimation error.

We can now generalise the conclusion at the end of Section V-A.1 by taking the imperfect channel estimation into account, and summarise that the MRT precoded system can be more robust to both reciprocity and channel estimation errors compared with the ZF precoded system.

*C. Implications*

In order to illustrate the implications that discussed in (52) and (57), we consider the results from Fig. 6, 7 and 8 to determine the proper values for the conditions in (52) and (57). Here let $M = 500$, $K = 20$ and $\rho_d = 20$ dB. We also consider a smaller value of the estimation error parameter, i.e., $\tau^2 = 0.01$. Since (52) and (57) are proportional to the term $A_I$, which is related to both amplitude and phase errors at both Tx/Rx RF frontends, let $\sigma_A^2 = \sigma_P^2$ to capture the aggregated variation of $A_I$. The other parameters have the same values of "High Level Reciprocity Error". We then derive $\text{SINR}_{\text{ZF/MRT}}$, i.e., the ratio of (47) to (46), and $\text{SINR}^{\text{err}}_{\text{ZF/MRT}}$, i.e., the ratio of (43) to (29), to demonstrate the output SINR comparison between MRT and ZF, corresponding to $\tilde{C}_I$ in (52) and $C_I$ in (57), respectively. It can be seen that the simulation-based results in Fig. 9 are tightly matched with analytical results of $\text{SINR}_{\text{ZF/MRT}}$ and $\text{SINR}^{\text{err}}_{\text{ZF/MRT}}$. We can also observe a close match between the analytical results of $\text{SINR}_{\text{ZF/MRT}}$ and $\text{SINR}^{\text{err}}_{\text{ZF/MRT}}$ with the asymptotic results $\tilde{C}_I$ and $C_I$ respectively. Furthermore, we can conclude from Fig. 9(a) that the performance preponderance of using ZF over MRT decreases precipitously when the level of the reciprocity error increases, and ends up with no gain compared to that of MRT. This conclusion holds in the case with the estimated channel as shown in Fig. 9(b). The match between our asymptotic result in (57) and the simulation results in (b) of Fig. 9, confirms that the gain of ZF is highly dependent on the quality of the channel reciprocity, and this gain can be independent of the estimation error especially with the severe reciprocity error introducing into the system, as discussed in Section IV. Along with the observation at the end of Section V-A.2, our results in this paper also indicate that MRT is more efficient compared to ZF in the high region of the reciprocity error, and in the relatively low region of the reciprocity error with the low transmit SNR or with the small ratio of $M/K$. However, we would like to note that further investigations, including the computational complexity of different precoding schemes, may be needed to provide a reliable comparison among different schemes.

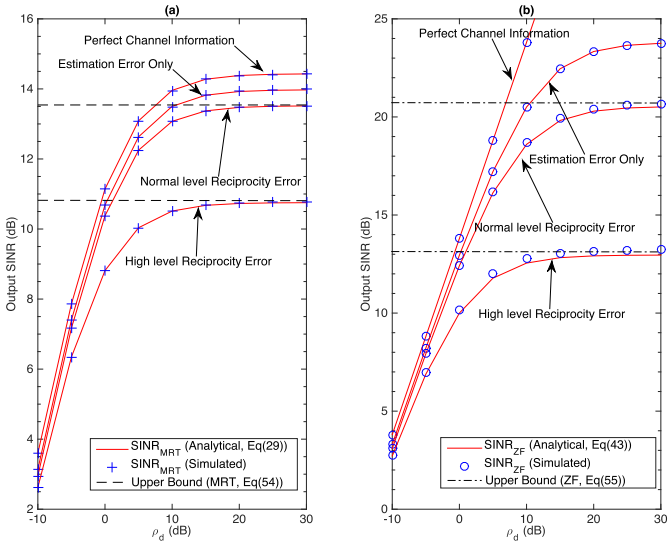

Fig. 8. Output SINR versus SNR in the presence of different levels of the channel reciprocity error and channel estimation error ($\tau^2 = 0.1$).

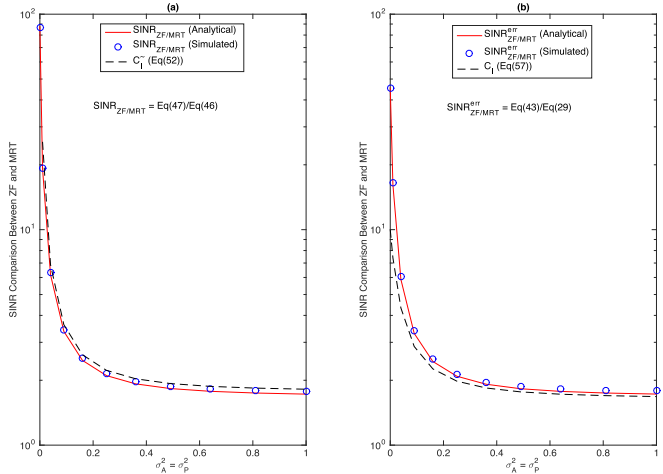

Fig. 9. Output SINR comparison of MRT and ZF.

## VI. CONCLUSIONS AND DISCUSSION

In this paper, we have analysed the impact of the channel reciprocity error caused by the RF mismatches, on the performance of linear precoding schemes such as MRT and ZF in TDD massive MU-MIMO systems with imperfect channel estimation. Considering the reciprocity errors as multiplicative uncertainties in the channel matrix with truncated Gaussian amplitude and phase errors, we have derived analytical expressions of the output SINR for MRT and ZF in the presence of the channel estimation error, and analysed the asymptotic behaviour of the system when the number of antennas at the BS is large. The perfect match has been found between the analytical and simulated results in the cases with the practical and asymptotically large values of the BS antennas, which verifies that our analytical results can be utilised to effectively evaluate the performance of the considered system.

Our analysis has taken into account the compound effect of both reciprocity error and estimation error on the system performance, which provides important engineering insights for practical TDD massive MIMO systems, such that: 1) the channel reciprocity error causes the error ceiling effect on the performance of massive MIMO systems even with the high SNR or large number of BS antennas, which can be held regardless of the existence of the channel estimation error; 2) ZF generally outperforms MRT in terms of the output SINR. However, MRT has better robustness to both reciprocity error and estimation error compared to ZF, thus can be more efficient than ZF in certain cases, e.g., in the high region of the reciprocity error, or in the low SNR regime. This would ultimately influence the choice of the precoding schemes for massive MIMO systems in the presence of the channel reciprocity error in practice.

Further investigations can be carried out by taking into account the computational complexity and energy efficiency of different precoding schemes, e.g., MRT, ZF, minimum mean square error (MMSE) or even the non-linear dirty paper coding, along with novel compensation techniques for massive MIMO systems suffer from the reciprocity error.

Our analysis can be generalised to large-scale fading scenarios by considering the effect of path loss or shadowing. For example, based on the analytical and simulated results of the output SINR versus different transmit SNR in this paper, one possible extension is to analyse the impact of distance-dependent path loss which can be simply reflected by the reduction of the transmit power.

## APPENDIX A
## PRELIMINARIES ON THE TRUNCATED GAUSSIAN DISTRIBUTION

A brief of the truncated Gaussian distribution is given here. Consider that $X$ is normally distributed with mean $\mu$ and variance $\sigma^2$, and lies within a truncated range $[a, b]$, where $-\infty < a < b < \infty$, then $X$ conditional on $a \leq X \leq b$ is treated to have truncated Gaussian distribution, which can be denoted by $X \sim \mathcal{N}_T(\mu, \sigma^2), X \in [a, b]$. For a given $x \in [a, b]$, the probability density function can be given as [28]

$$f(x, \mu, \sigma; a, b) = \frac{1}{\sigma Z}\phi\left(\frac{x-\mu}{\sigma}\right). \quad (59)$$

The revised expected value and variance conditioned on the truncated range $[a, b]$ can be written as

$$\mathbb{E}\{X\} = \mu + \frac{\phi(\alpha) - \phi(\beta)}{Z}\sigma, \quad (60)$$

$$\mathrm{var}(X) = \sigma^2\left[1 + \frac{\alpha\phi(\alpha) - \beta\phi(\beta)}{Z} - \left(\frac{\phi(\alpha) - \phi(\beta)}{Z}\right)^2\right], \quad (61)$$

where

$$\alpha = \frac{a-\mu}{\sigma}, \beta = \frac{b-\mu}{\sigma}, Z = \Phi(\beta) - \Phi(\alpha), \quad (62)$$

$$\phi(\cdot) = \frac{1}{\sqrt{2\pi}}\exp\left(-\frac{1}{2}(\cdot)^2\right), \quad (63)$$

$$\Phi(\cdot) = \frac{1}{2}\left(1 + \mathrm{erf}\left(\frac{1}{\sqrt{2}}(\cdot)\right)\right). \quad (64)$$

## APPENDIX B
## USEFUL EXTENSIONS

### A. Proof of Proposition 1

In general, given a random variable $x$ and its probability function $f(x)$, the expected value of a function of $x$ can be calculated by

$$\mathbb{E}\{g(x)\} = \int_{-\infty}^{\infty} f(x)g(x)\,\mathrm{d}x. \quad (65)$$

In this case, $f(x)$ is given by (59) with $x \in [a, b]$, and $g(x) = \exp(jx)$, thus we formulate $\mathbb{E}\{g(x)\}$ as

$$\mathbb{E}\{\exp(jx)\} = \int_a^b f(x, \mu, \sigma; a, b)\exp(jx)\,\mathrm{d}x$$

$$= \frac{1}{\sqrt{2\pi}\sigma Z}\int_a^b \exp\left(-\frac{1}{2}\left(\frac{x-\mu}{\sigma}\right)^2 + jx\right)\,\mathrm{d}x$$

$$= \frac{1}{\sqrt{2\pi}\sigma Z}\sqrt{\frac{\pi}{4\frac{1}{2\sigma^2}}}\exp\left(-\frac{\mu^2}{2\sigma^2} + \frac{\left(j+\frac{\mu}{\sigma^2}\right)^2}{4\frac{1}{2\sigma^2}}\right)$$

$$\times \mathrm{erf}\left(\sqrt{\frac{1}{2\sigma^2}}x - \frac{j+\frac{\mu}{\sigma^2}}{2\sqrt{\frac{1}{2\sigma^2}}}\right)\Bigg|_a^b$$

$$= \frac{1}{2Z}\exp\left(-\frac{\sigma^2}{2} + j\mu\right)$$

$$\times \left(\mathrm{erf}\left(\frac{\sqrt{2}}{2}\left(\frac{b-\mu}{\sigma}\right) - \frac{\sqrt{2}j\sigma}{2}\right)\right.$$

$$\left. - \mathrm{erf}\left(\frac{\sqrt{2}}{2}\left(\frac{a-\mu}{\sigma}\right) - \frac{\sqrt{2}j\sigma}{2}\right)\right). \quad (66)$$

By invoking (62) and (64) into (66), we arrive at the result in Proposition 1.

### B. Remarks

Based on the Proposition 1 that demonstrates a generic case for the given $x \sim \mathcal{N}_T(\mu, \sigma^2), x \in [a, b]$, useful remarks can be given as follows.

*Remark 1:* Let $\mu = 0$ in Proposition 1, then $\mathbb{E}\{\exp(jx)\}$ of $x \sim \mathcal{N}_T(0, \sigma^2), x \in [a, b]$ can be rewritten as

$$\mathbb{E}\{\exp(jx)\}$$
$$= \exp\left(-\frac{\sigma^2}{2}\right)\left(\frac{\mathrm{erf}\left(\frac{b}{\sqrt{2\sigma^2}} - j\frac{\sigma}{\sqrt{2}}\right) - \mathrm{erf}\left(\frac{a}{\sqrt{2\sigma^2}} - j\frac{\sigma}{\sqrt{2}}\right)}{\mathrm{erf}\left(\frac{b}{\sqrt{2\sigma^2}}\right) - \mathrm{erf}\left(\frac{a}{\sqrt{2\sigma^2}}\right)}\right). \quad (67)$$

*Remark 2:* Let $\mu = 0$ and $a = -b$ in Proposition 1, then $\mathbb{E}\{\exp(jx)\}$ of $x \sim \mathcal{N}_T(0, \sigma^2), x \in [-b, b]$ can be given as

$$\mathbb{E}\{\exp(jx)\} = \frac{\exp\left(-\frac{\sigma^2}{2}\right)}{\mathrm{erf}\left(\frac{b}{\sqrt{2\sigma^2}}\right)}\Re\left(\mathrm{erf}\left(\left(\frac{b}{\sqrt{2\sigma^2}}\right) \pm j\frac{\sigma}{\sqrt{2}}\right)\right). \quad (68)$$

## APPENDIX C
## USEFUL RESULTS

Recall (6), (8), (7) and (9), the random variables $A_{bt,i}$, $A_{br,i}$, $\varphi_{bt,i}$ and $\varphi_{br,i}$ can be regarded as truncated Gaussian variables, whose relative parameters can be given here. Specifically, based on the preliminaries on the truncated Gaussian distribution in Appendix A, the amplitude-error-related parameters can be expressed as

$$\alpha_t = \alpha_{bt,0} + \frac{\phi(\hat{a}_t) - \phi(\hat{b}_t)}{Z_t}\sigma_{bt}, \quad (69)$$

$$\sigma_t^2 = \sigma_{bt}^2\left(1 + \frac{\hat{a}_t\phi(\hat{a}_t) - \hat{b}_t\phi(\hat{b}_t)}{Z_t} - \left(\frac{\phi(\hat{a}_t) - \phi(\hat{b}_t)}{Z_t}\right)^2\right), \quad (70)$$

$$\alpha_r = \alpha_{br,0} + \frac{\phi(\hat{a}_r) - \phi(\hat{b}_r)}{Z_r}\sigma_{br}, \quad (71)$$

$$\sigma_r^2 = \sigma_{br}^2\left(1 + \frac{\hat{a}_r\phi(\hat{a}_r) - \hat{b}_r\phi(\hat{b}_r)}{Z_r} - \left(\frac{\phi(\hat{a}_r) - \phi(\hat{b}_r)}{Z_r}\right)^2\right), \quad (72)$$

where $\hat{a}_t = (a_t - \alpha_{bt,0})/\sigma_{bt}$, $\hat{b}_t = (b_t - \alpha_{bt,0})/\sigma_{bt}$, $\hat{a}_r = (a_r - \alpha_{br,0})/\sigma_{br}$, $\hat{b}_r = (b_r - \alpha_{br,0})/\sigma_{br}$, $Z_t = \Phi(\hat{b}_t) - \Phi(\hat{a}_t)$, and $Z_r = \Phi(\hat{b}_r) - \Phi(\hat{a}_r)$. The functions $\phi(\cdot)$ and $\Phi(\cdot)$ are given in (63) and (64) respectively.

Also, the phase-error-related functions $g_t$ and $g_r$ in (29) can be expressed by applying Proposition 1 and its proof in Appendix B, as given in the following.

$$g_t = \mathbb{E}\left\{\exp\left(j\varphi_{bt,i}\right)\right\} = \exp\left(-\frac{\sigma_{\varphi_t}^2}{2} + j\theta_{bt,0}\right)$$
$$\times \frac{\text{erf}\left(\left(\frac{\theta_{t,2}-\theta_{bt,0}}{\sqrt{2\sigma_{\varphi_t}^2}}\right) - j\frac{\sigma_{\varphi_t}}{\sqrt{2}}\right) - \text{erf}\left(\left(\frac{\theta_{t,1}-\theta_{bt,0}}{\sqrt{2\sigma_{\varphi_t}^2}}\right) - j\frac{\sigma_{\varphi_t}}{\sqrt{2}}\right)}{\text{erf}\left(\frac{\theta_{t,2}-\theta_{bt,0}}{\sqrt{2\sigma_{\varphi_t}^2}}\right) - \text{erf}\left(\frac{\theta_{t,1}-\theta_{bt,0}}{\sqrt{2\sigma_{\varphi_t}^2}}\right)}, \tag{73}$$

$$g_r = \mathbb{E}\left\{\exp\left(j\varphi_{br,i}\right)\right\} = \exp\left(-\frac{\sigma_{\varphi_r}^2}{2} + j\theta_{br,0}\right)$$
$$\times \frac{\text{erf}\left(\left(\frac{\theta_{r,2}-\theta_{br,0}}{\sqrt{2\sigma_{\varphi_r}^2}}\right) - j\frac{\sigma_{\varphi_r}}{\sqrt{2}}\right) - \text{erf}\left(\left(\frac{\theta_{r,1}-\theta_{br,0}}{\sqrt{2\sigma_{\varphi_r}^2}}\right) - j\frac{\sigma_{\varphi_r}}{\sqrt{2}}\right)}{\text{erf}\left(\frac{\theta_{r,2}-\theta_{br,0}}{\sqrt{2\sigma_{\varphi_r}^2}}\right) - \text{erf}\left(\frac{\theta_{r,1}-\theta_{br,0}}{\sqrt{2\sigma_{\varphi_r}^2}}\right)}. \tag{74}$$

## APPENDIX D
## PROOFS OF (26) AND (27)

In order to achieve the analytical expression of $\text{SINR}_{k,\text{mrt}}$ in Theorem 1, we calculate the normalisation parameter $\lambda_{\text{mrt}}$ in (23), the expectations of signal power scaling factor $P_{s,\text{mrt}}$ and the interference power scaling factor $P_{I,\text{mrt}}$ separately as following.

### A. $\lambda_{\text{mrt}}$

Consider the denominator inside of the square root sign in (23), we can have

$$\mathbb{E}\left\{\text{tr}\left(\mathbf{W}_{\text{mrt}}\mathbf{W}_{\text{mrt}}^H\right)\right\}$$
$$= (1-\tau^2)\mathbb{E}\left\{\text{tr}\left(\mathbf{H}_{br}^*\mathbf{H}^*\mathbf{H}^T\mathbf{H}_{br}\right)\right\} + \tau^2\mathbb{E}\left\{\text{tr}\left(\mathbf{V}^*\mathbf{V}^T\right)\right\} \tag{75}$$
$$= MK\left((1-\tau^2)(\alpha_r^2+\sigma_r^2)+\tau^2\right), \tag{76}$$

where (75) is conditioned on the independence between $\mathbf{H}$, $\mathbf{H}_{bt}$, $\mathbf{H}_{br}$ and $\mathbf{V}$.

### B. $\mathbb{E}\left\{P_{s,\text{mrt}}\right\}$

Considering the normalised symbol power of $s_k$ as mentioned in Section II, partial $\mathbb{E}\left\{P_{s,\text{mrt}}\right\}$, i.e., $\mathbb{E}\left\{|\mathbf{h}_k^T\mathbf{H}_{bt}\mathbf{w}_{k,\text{mrt}}|^2\right\}$, can be computed as

$$\mathbb{E}\left\{|\mathbf{h}_k^T\mathbf{H}_{bt}\mathbf{w}_{k,\text{mrt}}|^2\right\}$$
$$= \mathbb{E}\left\{|\mathbf{h}_k^T\mathbf{H}_{bt}(\sqrt{1-\tau^2}\mathbf{H}_{br}^*\mathbf{h}_k^* + \tau\mathbf{v}_k^*)|^2\right\}$$
$$= (1-\tau^2)\mathbb{E}\left\{|\mathbf{h}_k^T\mathbf{H}_{bt}\mathbf{H}_{br}^*\mathbf{h}_k^*|^2\right\} + \tau^2\mathbb{E}\left\{|\mathbf{h}_k^T\mathbf{H}_{bt}\mathbf{v}_k^*|^2\right\}, \tag{77}$$

where

$$\mathbb{E}\left\{|\mathbf{h}_k^T\mathbf{H}_{bt}\mathbf{H}_{br}^*\mathbf{h}_k^*|^2\right\}$$
$$= \mathbb{E}\left\{\sum_{i_1=1}^{M}|h_{i_1,k}|^2\left(h_{bt,i_1}h_{br,i_1}^*\right)\sum_{i_2=1}^{M}|h_{i_2,k}|^2\left(h_{bt,i_2}^*h_{br,i_2}\right)\right\} \tag{78}$$
$$= \sum_{i_1=1}^{M}\mathbb{E}\{|h_{i_1,k}|^4|h_{bt,i_1}|^2|h_{br,i_1}|^2$$
$$+ \sum_{i_2=1,i_2\neq i_1}^{M}|h_{i_1,k}|^2|h_{i_2,k}|^2 h_{bt,i_1}h_{br,i_1}^* h_{bt,i_2}^* h_{br,i_2}\} \tag{79}$$
$$= M\left(2(\alpha_t^2+\sigma_t^2)(\alpha_r^2+\sigma_r^2) + (M-1)\alpha_t^2\alpha_r^2|g_t|^2|g_r|^2\right), \tag{80}$$

and similarly,

$$\mathbb{E}\left\{|\mathbf{h}_k^T\mathbf{H}_{bt}\mathbf{v}_k^*|^2\right\} = M(\alpha_t^2+\sigma_t^2). \tag{81}$$

Next, by substituting (80) and (81) in (77), and invoking $\rho_d$, $\lambda_{\text{mrt}}$ in (23) and the completed (77), we have (26).

### C. $\mathbb{E}\left\{P_{I,\text{mrt}}\right\}$

By omitting the independent $s_i$ for different users with the normalised power, partial $\mathbb{E}\left\{P_{I,\text{mrt}}\right\}$ can be modified as $\mathbb{E}\left\{\left|\sum_{i=1,i\neq k}^{K}\mathbf{h}_k^T\mathbf{H}_{bt}\mathbf{w}_{i,\text{mrt}}\right|^2\right\}$, which is calculated as

$$\mathbb{E}\left\{\left|\sum_{i=1,i\neq k}^{K}\mathbf{h}_k^T\mathbf{H}_{bt}\mathbf{w}_{i,\text{mrt}}\right|^2\right\}$$
$$= \sum_{i=1,i\neq k}^{K}\left((1-\tau^2)\mathbb{E}\left\{|\mathbf{h}_k^T\mathbf{H}_{bt}\mathbf{H}_{br}^*\mathbf{h}_i^*|^2\right\}\right.$$
$$\left. +\tau^2\mathbb{E}\left\{|\mathbf{h}_k^T\mathbf{H}_{bt}\mathbf{v}_i^*|^2\right\}\right), \tag{82}$$

where

$$\mathbb{E}\left\{|\mathbf{h}_k^T\mathbf{H}_{bt}\mathbf{H}_{br}^*\mathbf{h}_i^*|^2\right\}$$
$$= \sum_{j_1=1}^{M}\mathbb{E}\{|h_{j_1,k}|^2|h_{j_1,i}|^2|h_{bt,j_1}|^2|h_{br,j_1}|^2$$
$$+ \sum_{j_2=1,j_2\neq j_1}^{M}h_{j_1,k}h_{j_2,k}^*h_{j_1,i}^*h_{j_2,i}h_{bt,j_1}h_{br,j_1}^*h_{bt,j_2}^*h_{br,j_2}\} \tag{83}$$
$$= M(\alpha_t^2+\sigma_t^2)(\alpha_r^2+\sigma_r^2). \tag{84}$$

And $\mathbb{E}\left\{|\mathbf{h}_k^T\mathbf{H}_{bt}\mathbf{v}_i^*|^2\right\}$ can be obtained as in (81). Then applying $\rho_d$, $\lambda_{\text{mrt}}$ in (23) and the completed result in (82), the expectation of $P_{I,\text{mrt}}$ can be given as in (27).

## APPENDIX E
## PROOFS OF PROPOSITION 3 AND 4

To formulate the signal and interference power as well as the output SINR in the case of ZF precoded system, the normalisation parameter $\lambda_{\text{zf}}$ can be calculated first.

## A. $\lambda_{\text{zf}}$

The same conditions as in Theorem 1 are applied when calculating $\lambda_{\text{zf}}$. The power constraint on $\mathbf{W}_{\text{zf}}$ can be extended as

$$\mathbb{E}\left\{\text{tr}\left(\mathbf{W}_{\text{zf}}\mathbf{W}_{\text{zf}}^H\right)\right\}$$

$$= \mathbb{E}\left\{\text{tr}\left(\hat{\mathbf{H}}_d^H\left(\hat{\mathbf{H}}_d\hat{\mathbf{H}}_d^H\right)^{-1}\left(\hat{\mathbf{H}}_d\hat{\mathbf{H}}_d^H\right)^{-1}\hat{\mathbf{H}}_d\right)\right\} \quad (85)$$

$$= \mathbb{E}\left\{\text{tr}\left(\left((\sqrt{1-\tau^2}\mathbf{H}^T\mathbf{H}_{br} + \tau\mathbf{V}^T)\right.\right.\right.$$
$$\left.\left.\left.\times (\sqrt{1-\tau^2}\mathbf{H}_{br}^*\mathbf{H}^* + \tau\mathbf{V}^*)\right)^{-1}\right)\right\} \quad (86)$$

$$\stackrel{(a)}{\approx} \mathbb{E}\left\{\text{tr}\left(\left((1-\tau^2)\mathbf{H}^T\mathbf{H}_{br}\mathbf{H}_{br}^*\mathbf{H}^* + \tau^2\mathbf{V}^T\mathbf{V}^*\right)^{-1}\right)\right\} \quad (87)$$

$$\stackrel{(b)}{\approx} \mathbb{E}\left\{\text{tr}\left(\left(\left(\frac{1-\tau^2}{M}\right)\mathbf{H}^T\text{tr}(\mathbf{H}_{br}\mathbf{H}_{br}^*)\mathbf{H}^* + \tau^2\mathbf{V}^T\mathbf{V}^*\right)^{-1}\right)\right\} \quad (88)$$

$$\stackrel{(c)}{=} \frac{1}{(1-\tau^2)(\alpha_r^2 + \sigma_r^2) + \tau^2}\mathbb{E}\left\{\text{tr}\left(\mathbf{W}_{\text{sum}}^{-1}\right)\right\} \quad (89)$$

$$\stackrel{(d)}{=} \frac{K}{(M-K)((1-\tau^2)(\alpha_r^2 + \sigma_r^2) + \tau^2)}, \quad (90)$$

where (a) is obtained due to the independence between the propagation channel $\mathbf{H}$ and the additive estimation error $\mathbf{V}$. Recall the assumption that $M$ is large, the term $\mathbf{H}^T\mathbf{H}_{br}\mathbf{H}_{br}^*\mathbf{H}^*$ tends to be diagonal, thus we can have (b) based on [29, eq. (14)]. Let $\mathbf{W}_{\text{sum}}$ represent the sum of $\mathbf{H}^T\mathbf{H}^*$ and $\mathbf{V}^T\mathbf{V}^*$, which are two independent Wishart matrices, then $\mathbf{W}_{\text{sum}}$ has a Wishart distribution whose the degree of freedom is the sum of the degrees of freedom of $\mathbf{H}^T\mathbf{H}^*$ and $\mathbf{V}^T\mathbf{V}^*$ [27], thus we have (c). And (d) can be achieved based on the random matrix theory as shown in [27]. Then we can arrive at the expression of $\lambda_{\text{zf}}$ in (38).

## B. $\mathbb{E}\left\{P_{s,\text{zf}}\right\}$

Consider the expectation of the signal power in (17) and recall $\mathbf{w}_{k,\text{zf}}$ as the $k$-th column of $\hat{\mathbf{H}}_d^H\left(\hat{\mathbf{H}}_d\hat{\mathbf{H}}_d^H\right)^{-1}$, we first compute the partial $\mathbb{E}\left\{P_{s,\text{zf}}\right\}$, i.e., $\mathbb{E}\left\{|\mathbf{h}_k^T\mathbf{H}_{bt}\mathbf{w}_{k,\text{zf}}|^2\right\}$, as follows,

$$\mathbb{E}\left\{|\mathbf{h}_k^T\mathbf{H}_{bt}\mathbf{w}_{k,\text{zf}}|^2\right\}$$
$$= \mathbb{E}\left\{|\mathbf{h}_k^T\mathbf{H}_{bt}[\hat{\mathbf{H}}_d^H(\hat{\mathbf{H}}_d\hat{\mathbf{H}}_d^H)^{-1}]_k|^2\right\}$$
$$= \mathbb{E}\left\{|\mathbf{h}_k^T\mathbf{H}_{bt}[(\sqrt{1-\tau^2}\mathbf{H}_{br}^*\mathbf{H}^* + \tau\mathbf{V}^*)\left((\sqrt{1-\tau^2}\mathbf{H}^T\mathbf{H}_{br} + \tau\mathbf{V}^T)(\sqrt{1-\tau^2}\mathbf{H}_{br}^*\mathbf{H}^* + \tau\mathbf{V}^*)\right)^{-1}]_k|^2\right\} \quad (91)$$

$$\approx \mathbb{E}\left\{|\mathbf{h}_k^T\mathbf{H}_{bt}[(\sqrt{1-\tau^2}\mathbf{H}_{br}^*\mathbf{H}^* + \tau\mathbf{V}^*)\right.$$
$$\left.\times \left((1-\tau^2)\mathbf{H}^T\mathbf{H}_{br}\mathbf{H}_{br}^*\mathbf{H}^* + \tau^2\mathbf{V}^T\mathbf{V}^*\right)^{-1}]_k|^2\right\}, \quad (92)$$

where $[\cdot]_k$ represents the k-th column of the matrix inside, and (92) can be achieved by applying (a) in deriving $\lambda_{\text{zf}}$. Consider the discussion following (90), when $M$ is large, $\left(\mathbf{H}^T\mathbf{H}_{br}\mathbf{H}_{br}^*\mathbf{H}^*\right)^{-1}$ becomes $(M/\text{tr}\left(\mathbf{H}_{br}\mathbf{H}_{br}^*\right))\left(\mathbf{H}^T\mathbf{H}^*\right)^{-1}$ asymptotically, and additionally, both $\mathbf{H}^T\mathbf{H}^*$ and $\mathbf{V}^T\mathbf{V}^*$ tend to be proportional to an identity matrix. Hence, we can approximate (92) as

$$\mathbb{E}\left\{|\mathbf{h}_k^T\mathbf{H}_{bt}\mathbf{w}_{k,\text{zf}}|^2\right\}$$
$$\approx \mathbb{E}\left\{|\mathbf{h}_k^T\mathbf{H}_{bt}[(\sqrt{1-\tau^2}\mathbf{H}_{br}^*\mathbf{H}^* + \tau\mathbf{V}^*)\right.$$
$$\left.\times \left((1-\tau^2)\text{tr}\left(\mathbf{H}_{br}\mathbf{H}_{br}^*\right) + \tau^2 M\right)^{-1}\mathbf{I}_K]_k|^2\right\}. \quad (93)$$

By using the technique in [29, eq. (14)], and considering the the independence between $\mathbf{H}$, $\mathbf{H}_{bt}$, $\mathbf{H}_{br}$ and $\mathbf{V}$, we have

$$\mathbb{E}\left\{|\mathbf{h}_k^T\mathbf{H}_{bt}\mathbf{w}_{k,\text{zf}}|^2\right\}$$
$$\approx \mathbb{E}\left\{|\sqrt{1-\tau^2}\left((1-\tau^2)\text{tr}\left(\mathbf{H}_{br}\mathbf{H}_{br}^*\right) + \tau^2 M\right)^{-1}\right.$$
$$\left.\times \mathbf{h}_k^T\mathbf{H}_{bt}\mathbf{H}_{br}^*\mathbf{h}_k^*|^2\right\} \quad (94)$$

$$\approx \frac{(1-\tau^2)\alpha_t^2\alpha_r^2|g_t|^2|g_r|^2}{\left((1-\tau^2)(\alpha_r^2 + \sigma_r^2) + \tau^2\right)^2}. \quad (95)$$

Therefore, by introducing (38) and (95) into $\mathbb{E}\left\{P_{s,\text{zf}}\right\}$, we can obtain (39) in Proposition 3.

## C. $\mathbb{E}\left\{P_{I,\text{zf}}\right\}$

Based on the complete result of $\mathbb{E}\left\{P_{s,\text{zf}}\right\}$ in Proposition 3 and $\lambda_{\text{zf}}$ in (38), the expected value of partial $P_{I,\text{zf}}$ omitting $s_i$ (i.e., $\mathbb{E}\{|\sum_{i=1,i\neq k}^K \sqrt{\rho_d}\lambda_{\text{zf}}\mathbf{h}_k^T\mathbf{H}_{bt}\mathbf{w}_{i,\text{zf}}|^2\}$) can be derived as

$$\mathbb{E}\left\{\left|\sum_{i=1,i\neq k}^K \sqrt{\rho_d}\lambda_{\text{zf}}\mathbf{h}_k^T\mathbf{H}_{bt}\mathbf{w}_{i,\text{zf}}\right|^2\right\}$$

$$= \sum_{i=1,i\neq k}^K \mathbb{E}\left\{|\sqrt{\rho_d}\lambda_{\text{zf}}\mathbf{h}_k^T\mathbf{H}_{bt}\mathbf{w}_{i,\text{zf}}|^2\right\} \quad (96)$$

$$\stackrel{(e)}{=} \rho_d\lambda_{\text{zf}}^2\left(\mathbb{E}\left\{\|\mathbf{h}_k^T\mathbf{H}_{bt}\mathbf{W}_{\text{zf}}\|^2\right\} - \mathbb{E}\left\{|\mathbf{h}_k^T\mathbf{H}_{bt}\mathbf{w}_{k,\text{zf}}|^2\right\}\right) \quad (97)$$

$$= \rho_d\lambda_{\text{zf}}^2 \mathbb{E}\left\{\|\mathbf{h}_k^T\mathbf{H}_{bt}\hat{\mathbf{H}}_d^H(\hat{\mathbf{H}}_d\hat{\mathbf{H}}_d^H)^{-1}\|^2\right\}$$
$$- \rho_d\lambda_{\text{zf}}^2 \mathbb{E}\left\{|\mathbf{h}_k^T\mathbf{H}_{bt}\mathbf{w}_{k,\text{zf}}|^2\right\} \quad (98)$$

$$\stackrel{(f)}{\approx} \frac{\rho_d\lambda_{\text{zf}}^2(K-1)}{M-K+1}\left(\frac{(\alpha_t^2 + \sigma_t^2)\left((1-\tau^2)\left(\alpha_r^2 + \sigma_r^2\right) + \tau^2\right)}{\left((1-\tau^2)(\alpha_r^2 + \sigma_r^2) + \tau^2\right)^2}\right.$$
$$\left. - \frac{(1-\tau^2)\alpha_t^2\alpha_r^2|g_t|^2|g_r|^2}{\left((1-\tau^2)(\alpha_r^2 + \sigma_r^2) + \tau^2\right)^2}\right) \quad (99)$$

$$\stackrel{(g)}{\approx} \frac{\rho_d(K-1)}{K}\left(\alpha_t^2 + \sigma_t^2 - \frac{(1-\tau^2)\alpha_t^2\alpha_r^2|g_t|^2|g_r|^2}{(1-\tau^2)(\alpha_r^2 + \sigma_r^2) + \tau^2}\right), \quad (100)$$

where (e) is due to the property of the ZF precoding scheme as in [10]. Based on Proposition 3 and [10], (f) is obtained by considering the diversity order of ZF, and (g) can be achieved under the assumption of the large ratio of $M/K$ in the massive MIMO system. To this end, we reach the approximated expression of $\mathbb{E}\{P_{I,\text{zf}}\}$ in Proposition 4.

## REFERENCES

[1] T. L. Marzetta, "Noncooperative cellular wireless with unlimited numbers of base station antennas," *IEEE Trans. Wireless Commun.*, vol. 9, no. 11, pp. 3590–3600, Nov. 2010.

[2] E. G. Larsson, O. Edfors, F. Tufvesson, and T. L. Marzetta, "Massive MIMO for next generation wireless systems," *IEEE Commun. Mag.*, vol. 52, no. 2, pp. 186–195, Feb. 2014.

[3] F. Boccardi, R. W. Heath, A. Lozano, T. L. Marzetta, and P. Popovski, "Five disruptive technology directions for 5G," *IEEE Commun. Mag.*, vol. 52, no. 2, pp. 74–80, Feb. 2014.

[4] Z. Gao, L. Dai, D. Mi, Z. Wang, M. Imran, and M. Shakir, "MmWave massive-MIMO-based wireless backhaul for the 5G ultra-dense network," *IEEE Wireless Commun. Mag.*, vol. 22, no. 5, pp. 13–21, Oct. 2015.

[5] A. Ijaz *et al.*, "Enabling massive IoT in 5G and beyond systems: PHY radio frame design considerations," *IEEE Access*, vol. 4, pp. 3322–3339, Sep. 2016.

[6] L. Lu, G. Y. Li, A. L. Swindlehurst, A. Ashikhmin, and R. Zhang, "An overview of massive MIMO: Benefits and challenges," *IEEE J. Sel. Topics Signal Process.*, vol. 8, no. 5, pp. 742–758, Oct. 2014.

[7] J. Hoydis, S. ten Brink, and M. Debbah, "Massive MIMO in the UL/DL of cellular networks: How many antennas do we need?" *IEEE J. Sel. Areas Commun.*, vol. 31, no. 2, pp. 160–171, Feb. 2013.

[8] H. Yang and T. L. Marzetta, "Performance of conjugate and zero-forcing beamforming in large-scale antenna systems," *IEEE J. Sel. Areas Commun.*, vol. 31, no. 2, pp. 172–179, Feb. 2013.

[9] F. Rusek *et al.*, "Scaling up MIMO: Opportunities and challenges with very large arrays," *IEEE Signal Process. Mag.*, vol. 30, no. 1, pp. 40–60, Jan. 2013.

[10] S. Wagner, R. Couillet, M. Debbah, and D. T. M. Slock, "Large system analysis of linear precoding in correlated MISO broadcast channels under limited feedback," *IEEE Trans. Inf. Theory*, vol. 58, no. 7, pp. 4509–4537, Jul. 2012.

[11] S. Vishwanath, N. Jindal, and A. Goldsmith, "Duality, achievable rates, and sum-rate capacity of Gaussian MIMO broadcast channels," *IEEE Trans. Inf. Theory*, vol. 49, no. 10, pp. 2658–2668, Oct. 2003.

[12] D. Mi, M. Dianati, S. Muhaidat, and Y. Chen, "A novel antenna selection scheme for spatially correlated massive MIMO uplinks with imperfect channel estimation," in *Proc. IEEE 81st Veh. Technol. Conf. (VTC)*, May 2015, pp. 1–6.

[13] J. Choi, D. J. Love, and P. Bidigare, "Downlink training techniques for FDD massive MIMO systems: Open-loop and closed-loop training with memory," *IEEE J. Sel. Topics Signal Process.*, vol. 8, no. 5, pp. 802–814, Oct. 2014.

[14] A. Tolli, M. Codreanu, and M. Juntti, "Compensation of non-reciprocal interference in adaptive MIMO-OFDM cellular systems," *IEEE Trans. Wireless Commun.*, vol. 6, no. 2, pp. 545–555, Feb. 2007.

[15] T. Schenk, *RF Imperfections in High-Rate Wireless Systems: Impact and Digital Compensation*. Dordrecht, The Netherlands: Springer, 2008. [Online]. Available: https://books.google.co.uk/books?id=nLzk11P15IAC

[16] E. Björnson, J. Hoydis, M. Kountouris, and M. Debbah, "Massive MIMO systems with non-ideal hardware: Energy efficiency, estimation, and capacity limits," *IEEE Trans. Inf. Theory*, vol. 60, no. 11, pp. 7112–7139, Nov. 2014.

[17] R. Rogalin *et al.*, "Scalable synchronization and reciprocity calibration for distributed multiuser MIMO," *IEEE Trans. Wireless Commun.*, vol. 13, no. 4, pp. 1815–1831, Apr. 2014.

[18] W. Zhang *et al.*, "Large-scale antenna systems with UL/DL hardware mismatch: Achievable rates analysis and calibration," *IEEE Trans. Commun.*, vol. 63, no. 4, pp. 1216–1229, Apr. 2015.

[19] D. Inserra and A. M. Tonello, "Characterization of hardware impairments in multiple antenna systems for DoA estimation," *J. Electr. Comput. Eng.*, vol. 2011, no. 18, pp. 1–10, Jan. 2011.

[20] *Channel Reciprocity Modeling and Performance Evaluation*, document TSG RAN WG1 59, R1-100426, Alcatel-Lucent, Boulogne-Billancourt, France, 3GPP, 2010.

[21] *Modelling of Channel Reciprocity Errors for TDD CoMP*, document TSG RAN WG1 64, R1-110804, Alcatel-Lucent, Boulogne-Billancourt, France, 3GPP, 2011.

[22] *Performance Study on Tx/Rx Mismatch in LTE TDD Dual-Layer Beamforming*, document TSG RAN WG1 57, R1-092550, 3GPP, Nokia, Nokia Siemens Netw., CATT, ZTE, 2009.

[23] A. Pitarokoilis, S. K. Mohammed, and E. G. Larsson, "Uplink performance of time-reversal MRC in massive MIMO systems subject to phase noise," *IEEE Trans. Wireless Commun.*, vol. 14, no. 2, pp. 711–723, Feb. 2015.

[24] D. Dobkin, *RF Engineering for Wireless Networks: Hardware, Antennas, and Propagation*. Amsterdam, The Netherlands: Elsevier, 2011.

[25] L. Zhang, A. U. Quddus, E. Katranaras, D. Wübben, Y. Qi, and R. Tafazolli, "Performance analysis and optimal cooperative cluster size for randomly distributed small cells under cloud RAN," *IEEE Access*, vol. 4, pp. 1925–1939, Sep. 2016.

[26] Y. Lim, C. Chae, and G. Caire, "Performance analysis of massive MIMO for cell-boundary users," *IEEE Trans. Wireless Commun.*, vol. 14, no. 12, pp. 6827–6842, Dec. 2015.

[27] A. M. Tulino and S. Verdu, *Random Matrix Theory and Wireless Communications*. Delft, The Netherlands: Now Publisher, 2004.

[28] N. L. Johnson, S. Kotz, and N. Balakrishnan, *Continuous Univariate Distributions*, 2nd ed. New York, NY, USA: Wiley, 1995.

[29] H. Wei, D. Wang, H. Zhu, J. Wang, S. Sun, and X. You, "Mutual coupling calibration for multiuser massive MIMO systems," *IEEE Trans. Wireless Commun.*, vol. 15, no. 1, pp. 606–619, Jan. 2016.

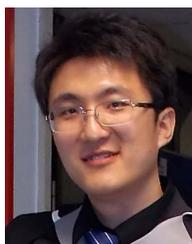

**De Mi** received the B.Eng. degree in information engineering from the Beijing Institute of Technology, Beijing, China, in 2011, and the M.Sc. degree in communications and signal processing from Imperial College, London, U.K., in 2012. He is currently pursuing the Ph.D. degree with the Institute for Communications Systems, University of Surrey, U.K. His research interests include massive MIMO and millimeter-wave communications.

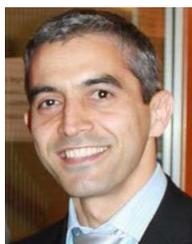

**Mehrdad Dianati** received the B.Sc. degree in electrical engineering from the Sharif University of Technology, Iran, the M.Sc. degree in electrical engineering from the K.N. Toosi University of Technology, Iran, and the Ph.D. degree in electrical and computer engineering from the University of Waterloo, Canada. He has been involved in a number of national and international projects as a Technical Coordinator and Work-Package Leader in recent years. He has nine years of industrial experience as a Senior Software/Hardware Developer and the Director of Research and Development. He is an Associate Editor of the IEEE TRANSACTIONS ON VEHICULAR TECHNOLOGY, the *IET Communications*, and the *Journal of Wireless Communications and Mobile* (Wiley).

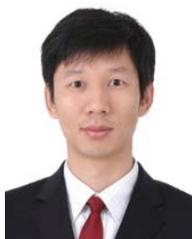

**Lei Zhang** received the B.Eng. degree in communication engineering and the M.Sc. degree in electromagnetic fields and microwave technology from the Northwestern Polytechnic University, Xi'an, China, in 2005 and 2008, respectively, and the Ph.D. degree from the Communications Research Group, University of Sheffield, U.K.. in 2011. He was a Research Engineer with Huawei Technologies, China. He is currently a Research Fellow in wireless communications with the Institute of Communications, University of Surrey, U.K. He holds over 10 international patents on wireless communications. His research interests include multi-antenna signal processing, air interface design, including waveform, frame structure, cloud radio access networks, and massive MIMO systems.

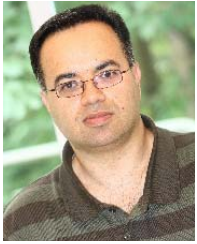 **Sami Muhaidat** (M'08-SM'11) received the Ph.D. degree in electrical and computer engineering from the University of Waterloo, Waterloo, ON, Canada, in 2006. From 2007 to 2008, he was an NSERC Post-Doctoral Fellow with the Department of Electrical and Computer Engineering, University of Toronto, Canada. From 2008 to 2012, he was an Assistant Professor with the School of Engineering Science, Simon Fraser University, BC, Canada. He is currently an Associate Professor with Khalifa University and also a Visiting Reader with the Faculty of Engineering, University of Surrey, U.K. He is also a Visiting Professor with the Department of Electrical and Computer Engineering, University of Western Ontario, Canada. His research focuses on advanced digital signal processing techniques for communications, cooperative communications, vehicular communications, MIMO, and machine learning. He has authored or co-authored over 130 technical papers on these topics. He currently serves as a Senior Editor of the IEEE COMMUNICATIONS LETTERS, an Editor of the IEEE TRANSACTIONS ON COMMUNICATIONS, and an Associate Editor of the IEEE TRANSACTIONS ON VEHICULAR TECHNOLOGY. He was a recipient of several scholarships during his undergraduate and graduate studies. He was also a recipient of the 2006 NSERC Post-doctoral Fellowship Competition.

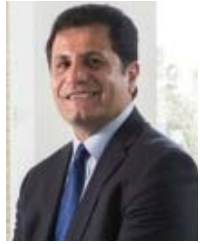 **Rahim Tafazolli** is currently a Professor and the Director of the Institute for Communication Systems and 5G Innovation Center, University of Surrey, U.K. He has authored over 500 research papers in refereed journals, international conferences and as Invited Speaker. He is the Editor of two books *Technologies for Wireless Future*, Vol. 1 (Wiley, 2004) and *Technologies for Wireless Future, Vol. 2* (Wiley, 2006). He also is Head of one of Europe's largest research groups. He was appointed as a Fellow of Wireless World Research Forum in 2011, in recognition of his personal contribution to the wireless world.